\documentclass[aps,twocolumn,preprintnumbers,amsmath,amssymb,superscriptaddress,floatfix,nofootinbib]{revtex4}
\usepackage{graphicx,color,dcolumn,booktabs,bm}
\usepackage{longtable,lscape}
\usepackage{txfonts}
\usepackage{overpic}
\usepackage{epsfig}
\usepackage{amssymb}
\usepackage{rotating}
\usepackage{epstopdf}
\usepackage{appendix}
\usepackage{indentfirst}
\usepackage{feynmf}   
\usepackage{slashed}  
\usepackage{cases}
\usepackage{color}
\usepackage{multirow}
\usepackage{graphicx,color,dcolumn,booktabs,bm}
\usepackage{cases}
\usepackage{array}
\def\c{\color{red}}

\graphicspath{{Figures/}} %

\usepackage[colorlinks, citecolor=blue,anchorcolor=red,menucolor=red, linkcolor=red,filecolor=red,runcolor=red,urlcolor=blue,frenchlinks=red]{hyperref}
\begin{document}
\title{Triply-heavy tetraquarks in an extended relativized quark model}

\author{Qi-Fang L\"u \footnote{Corresponding author.} } \email{lvqifang@hunnu.edu.cn} %
\affiliation{  Department
	of Physics, Hunan Normal University,  Changsha 410081, China }

\affiliation{ Synergetic Innovation
	Center for Quantum Effects and Applications (SICQEA), Changsha 410081,China}

\affiliation{  Key Laboratory of
	Low-Dimensional Quantum Structures and Quantum Control of Ministry
	of Education, Changsha 410081, China}

\author{Dian-Yong Chen} \email{chendy@seu.edu.cn} %
\affiliation {School of Physics, Southeast University, Nanjing 210094, China }

\author{Yu-Bing Dong} \email{dongyb@ihep.ac.cn} %
\affiliation{Institute of High Energy Physics, Chinese Academy of Sciences, Beijing 100049, China}

\affiliation{Theoretical Physics Center for Science Facilities (TPCSF), CAS, Beijing 100049, China}

\affiliation{School of Physical Sciences, University of Chinese Academy of Sciences, Beijing 101408, China}

\author{Elena Santopinto} \email{elena.santopinto@ge.infn.it} %
\affiliation{INFN, Sezione di Genova, Via Dodecaneso 33, 16146 Genova, Italy}

\begin{abstract}
In this paper, we adopt an extended relativized quark model to investigate the triply-heavy tetraquarks systematically. The mass spectra are obtained by solving the four-body relativized Hamiltonian including the Coulomb potential, confining potential, spin-spin interactions, and relativistic corrections. We find that all of the triply-heavy tetraquarks lie above the corresponding meson-meson thresholds, and thus no stable one exists. In particular, besides the spin-spin interactions, the Coulomb and confining potentials also contribute to the mass splittings in the $cb\bar c \bar q$ and $cb\bar b \bar q$ systems. Moreover, the whole mass spectra for triply heavy tetraquarks show quite similar patterns, which preserve the light flavor SU(3) symmetry and heavy quark symmetry well. Through the fall-apart mechanism, the triply-heavy tetraquarks may easily decay into the heavy quarkonium plus heavy-light mesons, which are good candidates for investigation in future experiments.
\end{abstract}

\maketitle

\section{Introduction}
\label{sec:intro}

In the past two decades, a large number of new hadron states have been observed experimentally, some of which are difficult to clarify as the conventional mesons or baryons. This significant experimental progress has attracted the extensive attention of theorists, and 
many studies have investigated the inner structures of these exotic states~\cite{Klempt:2007cp,Brambilla:2010cs,Hosaka:2016pey,Chen:2016qju,Lebed:2016hpi,Esposito:2016noz,Dong:2017gaw,Ali:2017jda,Guo:2017jvc,Olsen:2017bmm,Karliner:2017qhf,Liu:2019zoy,Brambilla:2019esw,Richard:2019cmi,Barabanov:2020jvn}. The various interpretations include conventional hadrons, loosely bound molecules, compact tetraquarks or pentaquarks, kinematic effects and so on. It is usually difficult to distinguish a compact tetraquark from a meson-meson molecule or conventional meson if they have the same quantum numbers. However, those with some heavy quarks, such as fully-and triply-heavy tetraquarks, are particularly interesting. For these systems, the molecular configurations are disfavored because no Goldstone boson exchange interaction appears. Also, they should lie far away from the scope of conventional meson spectra, which provides us with good opportunities to pin down the genuine compact tetraquarks.

Compared with other heavy tetraquarks, studies on triply-heavy tetraquarks are scarce in the literature. On using a nonrelativistic quark model, the authors did not find any bound triply-heavy tetraquarks~\cite{SilvestreBrac:1993ss}. The chromomagnetic interaction model and QCD sum rule have also been adopted to investigate these systems, and several weakly bound states are allowed within these frameworks~\cite{SilvestreBrac:1992mv,Cui:2006mp,Chen:2016ont,Jiang:2017tdc}. As a by-product of the mass spectra for doubly-heavy tetraquarks, the lattice QCD studies revealed that the lowest spin-1 $uc \bar b \bar b$ and $sc \bar b \bar b$ states were very near their corresponding meson-meson thresholds~\cite{Junnarkar:2018twb,Hudspith:2020tdf}. Further related discussions on triply-heavy tetraquarks can also be found in Refs.~\cite{Xing:2019wil,Liu:2019mxw}. As can be seen, these studies have not been able to reach a consistent conclusion and investigations on triply-heavy tetraquarks are far from sufficient.

In spite of the shortage of investigations, the triply-heavy tetraquarks may play an essential role as connecting bridges among different heavy tetraquark systems. As is known, the $X(6900)$ observed by the LHCb Collaboration is a good candidate for fully-heavy tetraquarks~\cite{Aaij:2020fnh}, and considerable attention has been paid to this structure~\cite{liu:2020eha,Wang:2020ols,Yang:2020rih,Jin:2020jfc,Lu:2020cns,Becchi:2020uvq,Chen:2020xwe,Wang:2020gmd,Albuquerque:2020hio,Sonnenschein:2020nwn,Giron:2020wpx,Maiani:2020pur,Richard:2020hdw,Chao:2020dml,Wang:2020wrp,Yang:2020atz,Maciula:2020wri,Karliner:2020dta,Wang:2020dlo,Dong:2020nwy,Ma:2020kwb,Feng:2020riv,Zhao:2020nwy,Gordillo:2020sgc,Faustov:2020qfm,Weng:2020jao,Zhang:2020xtb,Zhu:2020xni,Guo:2020pvt,Feng:2020qee,Cao:2020gul,Zhang:2020vpz,Zhu:2020snb,Gong:2020bmg,Dong:2020hxe,Wan:2020fsk,Wang:2020tpt,Yang:2020wkh,Liu:2020tqy,Huang:2020dci,Zhao:2020cfi,Goncalves:2021ytq}. Indeed, some studies on the fully charmed tetraquarks have been conducted for many years~\cite{Chao:1980dv,Ader:1981db,Iwasaki:1975pv,Zouzou:1986qh,Heller:1985cb,Lloyd:2003yc,Berezhnoy:2011xn,Karliner:2016zzc,Richard:2017vry,Anwar:2017toa,Wang:2017jtz,Debastiani:2017msn,Wu:2016vtq,Liu:2019zuc,Wang:2019rdo,Bedolla:2019zwg,Deng:2020iqw} before the observation $X(6900)$. The triply-heavy tetraquarks can be obtained by replacing one heavy antiquark with one light antiquark, and the strong interactions are also provided by the short range one-gluon-exchange as well as quark confinement. If the fully-heavy tetraquarks do exist, the triply-heavy tetraquarks may also form compactly. Moreover, the triply-heavy tetraquarks can be related to the doubly-heavy ones. For the doubly-heavy tetraquarks, the consensus has been reached that the isoscalar
$bb \bar u \bar d$ state should be stable against its strong and electromagnetic decays~\cite{Zouzou:1986qh,Heller:1986bt,Carlson:1987hh,SilvestreBrac:1993ss,Semay:1994ht,Pepin:1996id,Brink:1998as,Vijande:2003ki,Ebert:2007rn,Zhang:2007mu,Yang:2009zzp,Navarra:2007yw,Dias:2011mi,Du:2012wp,Bicudo:2015kna,Francis:2016hui,Bicudo:2017szl,Luo:2017eub,Karliner:2017qjm,Eichten:2017ffp,Mehen:2017nrh,Ali:2018ifm,Xing:2018bqt,Park:2018wjk,Junnarkar:2018twb,Deng:2018kly,Maiani:2019cwl,Leskovec:2019ioa,Liu:2019yye,Hernandez:2019eox,Yang:2019itm,Tang:2019nwv,Agaev:2020dba,Lu:2020rog,Hudspith:2020tdf,Cheng:2020wxa}, and the stabilities are closely associated with the mass ratios between heavy and light subsystems~\cite{Zouzou:1986qh,SilvestreBrac:1993ss,Hernandez:2019eox,Lu:2020rog}. Certainly, we can keep increasing the mass of one light antiquark in the doubly-heavy systems such that they turn into the triply-heavy ones. It is therefore interesting to investigate whether one or more stable triply-heavy tetraquaks exist. Another beneficial connection arises from the discrimination between singly-heavy tetraquarks and conventional heavy-light mesons. For instance, the nature of $D^*_{s0}(2317)$ is a long standing puzzle, which may have the $c \bar s$, $cq \bar s \bar q$, or even $c \bar s-cq \bar s \bar q$ flavor component~\cite{Chen:2016spr}. Since the light quark pair can create or annihilate dynamically, it is difficult to distinguish the different interpretations and pick out the genuine tetraquark states in these energy regions. However, the genuine tetraquarks can be easily recognized if the light quark pair is replaced with the $c\bar c$ or $b \bar b$ pair. With one heavy-quark pair excitation, the conventional heavy-light mesons become hidden charm and bottom triply-heavy tetraquarks, which are analogues of the singly-heavy tetraquarks. In brief, the studies on triply-heavy tetraquarks can increase our understanding of the heavy tetraquark systems, and a unified and systematic description of these tetraquark states is needed.

In previous works~\cite{Lu:2020rog,Lu:2020qmp,Lu:2020cns}, we extended the relativized quark model proposed by Godfrey and Isgur in order to investigate the singly-, doubly-, and fully-heavy tetraquark systems with the original model parameters. With such an extension, the tetraquaks and conventional mesons can be described in a uniform frame. Since the relativized potential gives a unified description of different flavor sectors and involves relativistic effects, it is believed to be more suitable to deal with both heavy-light and heavy-heavy quark interactions. Also, with regard to the triply heavy tetraquarks, we do not have to worry about the possible influence via light meson exchange interactions, which are absent from the relativized quark model. In this paper, we further study the mass spectra of the triply-heavy tetraquarks in the extended relativized quark model and discuss their possible strong decay behaviors, which may provide helpful information for future experimental searches.

This article is organized as follows. In section~\ref{model}, we present a review of the extended relativized quark model. The results and  discussions for the mass spectra and strong decay modes for the triply heavy tetraquarks are given in Section~\ref{results}. The last section provides a brief summary.

\section{Extended relativized quark model}{\label{model}}

To calculate the mass spectra of triply-heavy tetraquarks $Q_1 Q_2 \bar Q_3 \bar q_4$, we adopt an extension of the Godfrey and Isgur
meson relativized quark model, as developed recently~\cite{Lu:2020rog}.  This model is a natural extension of the relativized quark model for mesons constructed by Godfrey and Isgur to deal with the tetraquark states, in which the color interactions in
quark-gluon degrees of freedom are considered. The relevant Hamiltonian for a $Q_1 Q_2 \bar Q_3 \bar q_4$ state can be written as
\begin{equation}
H = H_0+\sum_{i<j}V_{ij}^{\rm oge}+\sum_{i<j}V_{ij}^{\rm conf}, \label{ham}
\end{equation}
where
\begin{equation}
H_0 = \sum_{i=1}^{4}(p_i^2+m_i^2)^{1/2}
\end{equation}
is the relativistic kinetic energy, $V_{ij}^{\rm oge}$ is the one-gluon-exchange pairwise potential together with the spin-spin interactions,
and $V_{ij}^{\rm conf}$ stands for the confinement potential.

In the present study, only the $S-$wave ground states are concentrated on, the spin-orbit and tensor interactions are therefore not included. The potential $V_{ij}^{\rm oge}$ can be expressed as
\begin{equation}
V_{ij}^{\rm{oge}} = \beta_{ij}^{1/2}\tilde G(r_{ij})\beta_{ij}^{1/2} + \delta_{ij}^{1/2+\epsilon_c}\frac{2\boldsymbol{S_i}
\cdot \boldsymbol{S_j}}{3m_im_j} \nabla^2\tilde G(r_{ij})\delta_{ij}^{1/2+\epsilon_c},
\end{equation}
with
\begin{equation}
\beta_{ij} = 1+\frac{p_{ij}^2}{(p_{ij}^2+m_i^2)^{1/2}(p_{ij}^2+m_j^2)^{1/2}},
\end{equation}
and
\begin{equation}
\delta_{ij} = \frac{m_im_j}{(p_{ij}^2+m_i^2)^{1/2}(p_{ij}^2+m_j^2)^{1/2}}.
\end{equation}
The $p_{ij}$ is the magnitude of the momentum of either of the quarks in the center-of-mass frame of the $ij$ quark subsystem, and the
$\epsilon_c$ is a free parameter reflecting the momentum dependence. Here, the smeared Coulomb potential $\tilde G(r_{ij})$ is
\begin{equation}
\tilde G(r_{ij}) = \boldsymbol F_i \cdot \boldsymbol F_j \sum_{k=1}^3\frac{\alpha_k}{r_{ij}}{\rm erf}(\tau_{kij}r_{ij}),
\end{equation}
with
\begin{equation}
\frac{1}{\tau_{kij}^2} = \frac{1}{\gamma_k^2}+\frac{1}{\sigma_{ij}^2},
\end{equation}
and
\begin{equation}
\sigma_{ij}^2 = \sigma_0^2
\left[\frac{1}{2}+\frac{1}{2}\left(\frac{4m_im_j}{(m_i+m_j)^2}\right)^4\right]+s^2\left(\frac{2m_im_j}{m_i+m_j}\right)^2,
\end{equation}
where the $\boldsymbol F_i \cdot \boldsymbol F_j$ stands for the color matrix and reads
\begin{equation}
\boldsymbol F_i = \left\{
\begin{aligned}
&\frac{\lambda_i}{2}~~~{\rm for~quarks,} \\
&-\frac{\lambda_i^*}{2}~{\rm for~antiquarks.} \\
\end{aligned}
\right.
\end{equation}
Similarly, the confining potential $V_{ij}^{\rm conf}$ can be expressed as
\begin{eqnarray}
V_{ij}^{\rm{conf}} &=& -\frac{3}{4}\boldsymbol F_i \cdot \boldsymbol F_j \nonumber \\ && \times
\left\{ br \left[\frac{e^{-\sigma_{ij}^2r^2}}{\sqrt{\pi}\sigma_{ij}r}+\left(1+\frac{1}{2\sigma_{ij}^2r^2}\right)
{\rm erf}(\sigma_{ij}r)\right]+c\right\}.\nonumber \\
\end{eqnarray}

All the parameters used here are taken from the original reference~\cite{Godfrey:1985xj} and collected in Table~\ref{para} for convenience. These parameters have been widely used for conventional mesons and tetraquarks with great success. The details of the relativized procedure can be found in Refs.~\cite{Godfrey:1985xj,Capstick:1986bm}.

\begin{table}[!htbp]
	\begin{center}
		\caption{ \label{para} Relevant parameters.}
		\begin{tabular*}{8.5cm}{@{\extracolsep{\fill}}*{5}{p{1.25cm}<{\centering}}}
			\hline\hline
			 $m_{u/d} (\rm MeV)$          &  $m_s (\rm MeV)$            & $m_c (\rm MeV)$       & $m_b (\rm MeV)$   & $\alpha_1$  \\
			220       &  419           & 1628    & 4977       & 0.25  \\\hline
			$\alpha_2$  &  $\alpha_3$   & $\gamma_1 (\rm GeV)$ & $\gamma_2 (\rm GeV)$  & $\gamma_3 (\rm GeV)$ \\
			0.15 & 0.20 & 1/2 & $\sqrt{10}/2$ & $\sqrt{1000}/2$ \\\hline
			$b (\rm GeV^2)$ & $c (\rm MeV)$     & $\sigma_0 (\rm GeV)$  &  $s$  & $\epsilon_c$ \\
			0.18 & $-253$  & 1.80  & 1.55  & $ -0.168$  \\
			\hline\hline
		\end{tabular*}
	\end{center}
\end{table}

The wave function for a $Q_1 Q_2 \bar Q_3 \bar q_4$ state consists of color, flavor, spin, and spatial parts. In the color space, two types of colorless states with certain permutation properties are employed
\begin{equation}
|\bar 3 3\rangle = |(Q_1 Q_2)^{\bar 3} (\bar Q_3 \bar q_4)^3\rangle,
\end{equation}
\begin{equation}
|6 \bar 6\rangle = |(Q_1 Q_2)^{6} (\bar Q_3 \bar q_4)^{\bar 6}\rangle,
\end{equation}
where the $|\bar 3 3\rangle$ and $|6 \bar 6 \rangle$ correspond to the antisymmetric and symmetric color wave functions under the exchange of $Q_1Q_2$ or $\bar Q_3 \bar q_4$, respectively. In the flavor space, the combinations $\{ c c \}$ and $\{b b \}$ are always symmetric, while the $c$, $b$ and $q$ are treated as different particles without symmetry constraints.

For the spin part, the six basic spin states are
\begin{equation}
\chi^{00}_0 = |(Q_1 Q_2)_0 (\bar Q_3 \bar q_4)_0\rangle_0,
\end{equation}
\begin{equation}
\chi^{11}_0 = |(Q_1 Q_2)_1 (\bar Q_3 \bar q_4)_1\rangle_0,
\end{equation}
\begin{equation}
\chi^{01}_1 = |(Q_1 Q_2)_0 (\bar Q_3 \bar q_4)_1\rangle_1,
\end{equation}
\begin{equation}
\chi^{10}_1 = |(Q_1 Q_2)_1 (\bar Q_3 \bar q_4)_0\rangle_1,
\end{equation}
\begin{equation}
\chi^{11}_1 = |(Q_1 Q_2)_1 (\bar Q_3 \bar q_4)_1\rangle_1,
\end{equation}
\begin{equation}
\chi^{11}_2 = |(Q_1 Q_2)_1 (\bar Q_3 \bar q_4)_1\rangle_2,
\end{equation}
where $(Q_1 Q_2)_0$ and $(\bar Q_3 \bar q_4)_0$ are antisymmetric, and the
$(Q_1 Q_2)_1$ and $(\bar Q_3 \bar q_4)_1$ are symmetric for the two fermions under permutations.  In the notation $\chi^{S_{12}S_{34}}_S$, the $S_{12}$, $S_{34}$, and $S$ stand for the spin of two quarks, spin of two antiquarks, and total spin, respectively. The relevant matrix elements of the color and spin parts are identical for various tetraquarks, and can be found in Ref.~\cite{Lu:2020rog}.

\begin{figure*}[!htbp]
\centering
\includegraphics[scale=0.7]{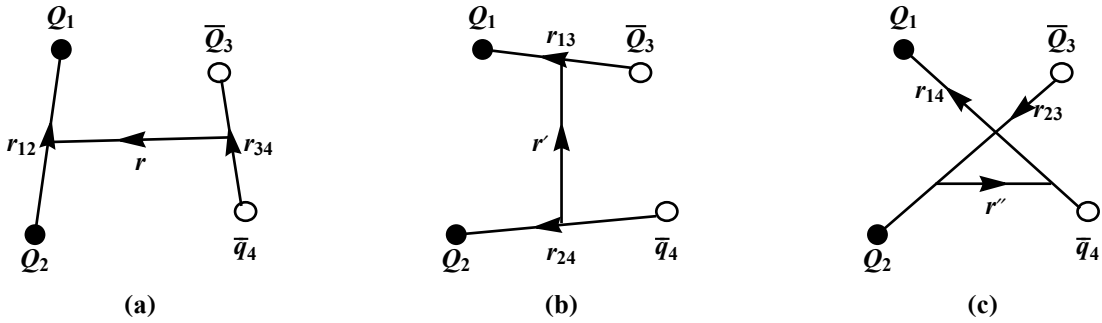}
\vspace{0.0cm} \caption{The $Q_1Q_2 \bar Q_3 \bar q_4$ tetraquark state in Jacobi coordinates. (a) combination $(Q_1Q_2)(\bar Q_3 \bar q_4)$ with relative coordinates $\boldsymbol r_{12}$, $\boldsymbol r_{34}$, and $\boldsymbol r$, (b) combination $(Q_1 \bar Q_3)(Q_2 \bar q_4)$ with relative coordinates $\boldsymbol r_{13}$, $\boldsymbol r_{24}$, and $\boldsymbol r^\prime$, (c) combination $(Q_1 \bar q_4)(Q_2 \bar Q_3)$ with relative coordinates $\boldsymbol r_{14}$, $\boldsymbol r_{23}$, and $\boldsymbol r^{\prime \prime}$.}
\label{jacobi}
\end{figure*}

In the spatial space, the Jacobi coordinates are presented in Fig.~\ref{jacobi}. For a $Q_1 Q_2 \bar Q_3 \bar q_4$ state, we can define
\begin{equation}
\boldsymbol r_{12}=  \boldsymbol r_1 - \boldsymbol r_2,
\end{equation}
\begin{equation}
\boldsymbol r_{34}=  \boldsymbol r_3 - \boldsymbol r_4,
\end{equation}
\begin{equation}
\boldsymbol r = \frac{m_1 \boldsymbol r_1 + m_2 \boldsymbol r_2}{m_1+m_2} - \frac{m_3 \boldsymbol r_3 + m_4 \boldsymbol r_4}{m_3+m_4},
\end{equation}
and
\begin{equation}
\boldsymbol R = \frac{m_1 \boldsymbol r_1 + m_2 \boldsymbol r_2 + m_3 \boldsymbol r_3 + m_4 \boldsymbol r_4}{m_1+m_2+m_3+m_4}.
\end{equation}
Thus, any other relative coordinates can be expressed in terms of $\boldsymbol r_{12}$, $\boldsymbol r_{34}$, and $\boldsymbol r$~\cite{Lu:2020rog}. For an $S-$wave state, a set of Gaussian
functions is adopted in order to approach its realistic spatial wave function~\cite{Hiyama:2003cu}
\begin{equation}
\Psi(\boldsymbol r_{12},\boldsymbol r_{34},\boldsymbol r) = \sum_{n_{12},n_{34},n} C_{n_{12}n_{34}n}
\psi_{n_{12}}(\boldsymbol r_{12}) \psi_{n_{34}}(\boldsymbol r_{34}) \psi_n(\boldsymbol r),
\end{equation}
where $C_{n_{12}n_{34}n}$ are the expansion coefficients. The $\psi_{n_{12}}(\boldsymbol r_{12}) \psi_{n_{34}}(\boldsymbol r_{34})
\psi_n(\boldsymbol r)$ is the position representation of the basis $|n_{12}n_{34}n\rangle$, where
\begin{equation}
\psi_n(\boldsymbol r) = \frac{2^{7/4}\nu_n^{3/4}}{\pi^{1/4}} e^{-\nu_n r^2} Y_{00}(\hat{\boldsymbol r}) =
\Bigg(\frac{2 \nu_n}{\pi} \Bigg )^{3/4} e^{-\nu_n r^2},
\end{equation}
\begin{equation}
\nu_n = \frac{1}{r_1^2a^{2(n-1)}},~~~~ (n=1, 2, ... , N_{max}).
\end{equation}
The $\psi_{n_{12}}(\boldsymbol r_{12})$ and  $\psi_{n_{34}}(\boldsymbol r_{34})$
can be written in a similar way, and the momentum representation the basis $ |n_{12}n_{34}n\rangle$ can be obtained through the Fourier transformation. It should be stressed that our final results are independent of geometric Gaussian size parameters $r_1$, $a$, and $N_{max}$ when
adequate bases are chosen~\cite{Hiyama:2003cu}, and the numerical stability is sufficient for quark model predictions~\cite{Lu:2020rog}.

According to the Pauli exclusion principle, the total wave function of a tetraquark should be antisymmetric; all possible configurations for the $Q_1Q_2 \bar Q_3 \bar q_4$ systems are presented in Table~\ref{configuration}. With the full wave functions, all of the matrix elements invovled in the Hamiltonian can be obtained straightforwardly. The masses without a mixing mechanism can therefore be calculated by solving the generalized eigenvalue problem
\begin{equation}
\sum_{j=1}^{N_{max}^3}(H_{ij}-EN_{ij})C_j=0,~~~~ (i=1, 2, ... , N_{max}^3),
\end{equation}
where the $H_{ij}$ are the matrix elements of the Hamiltonian, $N_{ij}$ are the overlap matrix elements of the Gaussian functions arising from their nonorthogonality, $E$ stands for the mass, and $C_j$ is the eigenvector corresponding to the coefficients $C_{n_{12}n_{34}n}$ of the spatial wave function. Moreover, for a given $Q_1Q_2 \bar Q_3 \bar q_4$ system, different configurations with the same $J^P$ can mix with each other. The mixing effects are taken into account here, and the final mass spectra and wave functions are obtained by diagonalizing the mass matrix of these configurations.

\begin{table*}[!htbp]
\begin{center}
\caption{ \label{configuration} All possible configurations for the $Q_1Q_2\bar Q_3 \bar q_4$ systems. The notation $n$ stands for the up or down quark, and the $s$ represents the strange quark. The subscripts and superscripts are the spin and color types, respectively. The braces $\{ ~\}$ stand for the symmetric flavor wave functions, and the parentheses $(~)$  are adopted for the subsystems without permutation symmetries.}
\begin{tabular*}{18cm}{@{\extracolsep{\fill}}*{5}{p{2.5cm}<{\centering}}}
\hline\hline
System & $J^P$         &  \multicolumn{3}{c}{Configuration} \\\hline
$\{ c c\} (\bar c \bar n)$ & $0^+$       &  $|\{cc\}^{\bar 3}_1 (\bar c \bar n)^3_1\rangle_0$     &  $|\{cc\}^6_0 (\bar c \bar n)^{\bar 6}_0\rangle_0$     & $\cdots$  \\
& $1^+$       &  $|\{cc\}^{\bar 3}_1 (\bar c \bar n)^3_0\rangle_1$  &  $|\{cc\}^{\bar 3}_1 (\bar c \bar n)^3_1\rangle_1$     & $|\{cc\}^6_0 (\bar c \bar n)^{\bar 6}_1\rangle_1$   \\
& $2^+$       &  $|\{cc\}^{\bar 3}_1 (\bar c \bar n)^3_1\rangle_2$     &  $\cdots$   & $\cdots$    \\

$\{ c c\} (\bar c \bar s)$ & $0^+$       &  $|\{cc\}^{\bar 3}_1 (\bar c \bar s)^3_1\rangle_0$     &  $|\{cc\}^6_0 (\bar c \bar s)^{\bar 6}_0\rangle_0$     & $\cdots$  \\
& $1^+$       &  $|\{cc\}^{\bar 3}_1 (\bar c \bar s)^3_0\rangle_1$  &  $|\{cc\}^{\bar 3}_1 (\bar c \bar s)^3_1\rangle_1$     & $|\{cc\}^6_0 (\bar c \bar s)^{\bar 6}_1\rangle_1$   \\
& $2^+$       &  $|\{cc\}^{\bar 3}_1 (\bar c \bar s)^3_1\rangle_2$     &  $\cdots$   & $\cdots$    \\

$\{ c c\} (\bar b \bar n)$ & $0^+$       &  $|\{cc\}^{\bar 3}_1 (\bar b \bar n)^3_1\rangle_0$     &  $|\{cc\}^6_0 (\bar b \bar n)^{\bar 6}_0\rangle_0$     & $\cdots$  \\
& $1^+$       &  $|\{cc\}^{\bar 3}_1 (\bar b \bar n)^3_0\rangle_1$  &  $|\{cc\}^{\bar 3}_1 (\bar b \bar n)^3_1\rangle_1$     & $|\{cc\}^6_0 (\bar b \bar n)^{\bar 6}_1\rangle_1$   \\
& $2^+$       &  $|\{cc\}^{\bar 3}_1 (\bar b \bar n)^3_1\rangle_2$     &  $\cdots$   & $\cdots$    \\

$\{ c c\} (\bar b \bar s)$ & $0^+$       &  $|\{cc\}^{\bar 3}_1 (\bar b \bar s)^3_1\rangle_0$     &  $|\{cc\}^6_0 (\bar b \bar s)^{\bar 6}_0\rangle_0$     & $\cdots$  \\
& $1^+$       &  $|\{cc\}^{\bar 3}_1 (\bar b \bar s)^3_0\rangle_1$  &  $|\{cc\}^{\bar 3}_1 (\bar b \bar s)^3_1\rangle_1$     & $|\{cc\}^6_0 (\bar b \bar s)^{\bar 6}_1\rangle_1$   \\
& $2^+$       &  $|\{cc\}^{\bar 3}_1 (\bar b \bar s)^3_1\rangle_2$     &  $\cdots$   & $\cdots$    \\\hline

$\{ bb\} (\bar c \bar n)$ & $0^+$       &  $|\{bb\}^{\bar 3}_1 (\bar c \bar n)^3_1\rangle_0$     &  $|\{bb\}^6_0 (\bar c \bar n)^{\bar 6}_0\rangle_0$     & $\cdots$  \\
& $1^+$       &  $|\{bb\}^{\bar 3}_1 (\bar c \bar n)^3_0\rangle_1$  &  $|\{bb\}^{\bar 3}_1 (\bar c \bar n)^3_1\rangle_1$     & $|\{bb\}^6_0 (\bar c \bar n)^{\bar 6}_1\rangle_1$   \\
& $2^+$       &  $|\{bb\}^{\bar 3}_1 (\bar c \bar n)^3_1\rangle_2$     &  $\cdots$   & $\cdots$    \\

$\{ bb\} (\bar c \bar s)$ & $0^+$       &  $|\{bb\}^{\bar 3}_1 (\bar c \bar s)^3_1\rangle_0$     &  $|\{bb\}^6_0 (\bar c \bar s)^{\bar 6}_0\rangle_0$     & $\cdots$  \\
& $1^+$       &  $|\{bb\}^{\bar 3}_1 (\bar c \bar s)^3_0\rangle_1$  &  $|\{bb\}^{\bar 3}_1 (\bar c \bar s)^3_1\rangle_1$     & $|\{bb\}^6_0 (\bar c \bar s)^{\bar 6}_1\rangle_1$   \\
& $2^+$       &  $|\{bb\}^{\bar 3}_1 (\bar c \bar s)^3_1\rangle_2$     &  $\cdots$   & $\cdots$    \\

$\{ bb\} (\bar b \bar n)$ & $0^+$       &  $|\{bb\}^{\bar 3}_1 (\bar b \bar n)^3_1\rangle_0$     &  $|\{bb\}^6_0 (\bar b \bar n)^{\bar 6}_0\rangle_0$     & $\cdots$  \\
& $1^+$       &  $|\{bb\}^{\bar 3}_1 (\bar b \bar n)^3_0\rangle_1$  &  $|\{bb\}^{\bar 3}_1 (\bar b \bar n)^3_1\rangle_1$     & $|\{bb\}^6_0 (\bar b \bar n)^{\bar 6}_1\rangle_1$   \\
& $2^+$       &  $|\{bb\}^{\bar 3}_1 (\bar b \bar n)^3_1\rangle_2$     &  $\cdots$   & $\cdots$    \\

$\{ bb\} (\bar b \bar s)$ & $0^+$       &  $|\{bb\}^{\bar 3}_1 (\bar b \bar s)^3_1\rangle_0$     &  $|\{bb\}^6_0 (\bar b \bar s)^{\bar 6}_0\rangle_0$     & $\cdots$  \\
& $1^+$       &  $|\{bb\}^{\bar 3}_1 (\bar b \bar s)^3_0\rangle_1$  &  $|\{bb\}^{\bar 3}_1 (\bar b \bar s)^3_1\rangle_1$     & $|\{bb\}^6_0 (\bar b \bar s)^{\bar 6}_1\rangle_1$   \\
& $2^+$       &  $|\{bb\}^{\bar 3}_1 (\bar b \bar s)^3_1\rangle_2$     &  $\cdots$   & $\cdots$    \\\hline

$(cb)(\bar c \bar n)$ & $ 0^+$  &  $|(cb)^{\bar 3}_0 (\bar c \bar n)^3_0 \rangle_0$  & $|(cb)^{\bar 3}_1 (\bar c \bar n)^3_1\rangle_0$   &  $|(cb)^6_0 (\bar c \bar n)^{\bar 6}_0\rangle_0$  \\
& & $|(cb)^6_1 (\bar c \bar n)^{\bar 6}_1\rangle_0$    & $\cdots$ & $\cdots$  \\
& $1^+$       &  $|(cb)^{\bar 3}_0 (\bar c \bar n)^3_1\rangle_1$ & $|(cb)^{\bar 3}_1 (\bar c \bar n)^3_0\rangle_1$  &$|(cb)^{\bar 3}_1 (\bar c \bar n)^3_1\rangle_1$ \\
& & $|(cb)^6_0 (\bar c \bar n)^{\bar 6}_1\rangle_1$  & $|(cb)^6_1 (\bar c \bar n)^{\bar 6}_0\rangle_1$ &  $|(cb)^6_1 (\bar c \bar n)^{\bar 6}_1\rangle_1$    \\
& $2^+$       &  $|(cb)^{\bar 3}_1 (\bar c \bar n)^3_1\rangle_2$     &  $|(cb)^6_1 (\bar c \bar n)^{\bar 6}_1\rangle_2$   &  $\cdots$    \\

$(cb)(\bar c \bar s)$ & $ 0^+$  &  $|(cb)^{\bar 3}_0 (\bar c \bar s)^3_0 \rangle_0$  & $|(cb)^{\bar 3}_1 (\bar c \bar s)^3_1\rangle_0$   &  $|(cb)^6_0 (\bar c \bar s)^{\bar 6}_0\rangle_0$  \\
& & $|(cb)^6_1 (\bar c \bar s)^{\bar 6}_1\rangle_0$    & $\cdots$ & $\cdots$  \\
& $1^+$       &  $|(cb)^{\bar 3}_0 (\bar c \bar s)^3_1\rangle_1$ & $|(cb)^{\bar 3}_1 (\bar c \bar s)^3_0\rangle_1$  &$|(cb)^{\bar 3}_1 (\bar c \bar s)^3_1\rangle_1$ \\
& & $|(cb)^6_0 (\bar c \bar s)^{\bar 6}_1\rangle_1$  & $|(cb)^6_1 (\bar c \bar s)^{\bar 6}_0\rangle_1$ &  $|(cb)^6_1 (\bar c \bar s)^{\bar 6}_1\rangle_1$    \\
& $2^+$       &  $|(cb)^{\bar 3}_1 (\bar c \bar s)^3_1\rangle_2$     &  $|(cb)^6_1 (\bar c \bar s)^{\bar 6}_1\rangle_2$   &  $\cdots$    \\

$(cb)(\bar b \bar n)$ & $ 0^+$  &  $|(cb)^{\bar 3}_0 (\bar b \bar n)^3_0 \rangle_0$  & $|(cb)^{\bar 3}_1 (\bar b \bar n)^3_1\rangle_0$   &  $|(cb)^6_0 (\bar b \bar n)^{\bar 6}_0\rangle_0$  \\
& & $|(cb)^6_1 (\bar b \bar n)^{\bar 6}_1\rangle_0$    & $\cdots$ & $\cdots$  \\
& $1^+$       &  $|(cb)^{\bar 3}_0 (\bar b \bar n)^3_1\rangle_1$ & $|(cb)^{\bar 3}_1 (\bar b \bar n)^3_0\rangle_1$  &$|(cb)^{\bar 3}_1 (\bar b \bar n)^3_1\rangle_1$ \\
& & $|(cb)^6_0 (\bar b \bar n)^{\bar 6}_1\rangle_1$  & $|(cb)^6_1 (\bar b \bar n)^{\bar 6}_0\rangle_1$ &  $|(cb)^6_1 (\bar b \bar n)^{\bar 6}_1\rangle_1$    \\
& $2^+$       &  $|(cb)^{\bar 3}_1 (\bar b \bar n)^3_1\rangle_2$     &  $|(cb)^6_1 (\bar b \bar n)^{\bar 6}_1\rangle_2$   &  $\cdots$    \\

$(cb)(\bar b \bar s)$ & $ 0^+$  &  $|(cb)^{\bar 3}_0 (\bar b \bar s)^3_0 \rangle_0$  & $|(cb)^{\bar 3}_1 (\bar b \bar s)^3_1\rangle_0$   &  $|(cb)^6_0 (\bar b \bar s)^{\bar 6}_0\rangle_0$  \\
& & $|(cb)^6_1 (\bar b \bar s)^{\bar 6}_1\rangle_0$    & $\cdots$ & $\cdots$  \\
& $1^+$       &  $|(cb)^{\bar 3}_0 (\bar b \bar s)^3_1\rangle_1$ & $|(cb)^{\bar 3}_1 (\bar b \bar s)^3_0\rangle_1$  &$|(cb)^{\bar 3}_1 (\bar b \bar s)^3_1\rangle_1$ \\
& & $|(cb)^6_0 (\bar b \bar s)^{\bar 6}_1\rangle_1$  & $|(cb)^6_1 (\bar b \bar s)^{\bar 6}_0\rangle_1$ &  $|(cb)^6_1 (\bar b \bar s)^{\bar 6}_1\rangle_1$    \\
& $2^+$       &  $|(cb)^{\bar 3}_1 (\bar b \bar s)^3_1\rangle_2$     &  $|(cb)^6_1 (\bar b \bar s)^{\bar 6}_1\rangle_2$   &  $\cdots$    \\\hline\hline

\end{tabular*}
\end{center}
\end{table*}

\section{Mass spectra and strong decay modes}{\label{results}}

The Gaussian expansion method has been wildly used for the few-body systems and has achieved great success. Empirically, stable results for $S-$ wave states can be achieved within small numbers of bases in the nonrelativistic quark model. In the relativized quark model, the number of bases should be large enough to guarantee completeness, otherwise the matrix elements of the terms $V_{ij}^{oge}$ will be meaningless. For the meson spectra, a dozen bases are adequate, while about one hundred bases are needed for the
baryon spectra~\cite{Godfrey:1985xj,Capstick:1986bm}. One can expect that several hundred or one thousand Gaussian functions are
suitable for calculating the tetraquark spectra. Indeed, we have investigated the singly-, doubly-, and fully-heavy tetraquark systems by using one thousand Gaussian bases in order to obtain their mass spectra with good convergency and high precision. In the present paper, we adopt $N^3_{max} = 10^3$ Gaussian bases to calculate the mass spectra of $S-$wave $Q_1Q_2 \bar Q_3 \bar q_4$ tetraquarks.  With these large bases, the numerical results are stable enough for quark model predictions. According to the flavor wave functions, these tetraquarks can be simply divided into three classes: (1) $cc\bar c \bar q$ and $cc\bar b \bar q$; (2) $bb\bar c \bar q$ and $bb\bar b \bar q$; and (3) $cb\bar c \bar q$ and $cb\bar b \bar q$. Here, we examine the mass spectra and strong decay modes for these systems successively, and provide some discussions on triply-heavy tetraquark systems.

\subsection{The $cc\bar c \bar q$ and $cc\bar b \bar q$ systems}

The predicted mass spectra for $cc\bar c \bar q$ and $cc\bar b \bar q$ systems are listed in Table~\ref{mass1}. The masses of the hidden charm $cc\bar c \bar q$ states lie in the range of $5400 \sim 5599$ MeV. A $cc\bar c \bar q$ state has the same quantum numbers as the $P-$wave conventional charmed or charmed-strange meson, and one more $c\bar c$ pair exists in the flavor wave function. However, the predicted masses of these $cc\bar c \bar q$ tetraquarks are much greater than $D$ and $D_s$ spectra, which indicates that the $cc\bar c \bar q$ tetraquarks are unlikely to mix with the conventional mesons. As in the case of conventional mesons, the mass splitting for these tetraquarks arises from the the spin-spin interactions. For the $cc\bar c \bar n$ and $cc\bar c \bar s$ states, their mass splittings are 133 and 123 MeV, respectively, which are comparable to those of $D$ and $D_s$ mesons.

\begin{table*}[htp]
\begin{center}
\caption{\label{mass1} Predicted mass spectra for the $cc\bar c \bar q$ and $cc\bar b \bar q$ systems.}
\begin{tabular*}{18cm}{@{\extracolsep{\fill}}p{1cm}<{\centering}p{1.7cm}<{\centering}p{4.5cm}<{\centering}p{1.8cm}<{\centering}p{4.5cm}<{\centering}}
\hline\hline
 $J^{P}$  & Configuration                                             & $\langle H\rangle$ (MeV) & Mass (MeV)  & Eigenvector\\\hline

 $0^{+}$  & $|\{cc\}^{\bar 3}_1 (\bar c \bar n)^3_1 \rangle_0$    & \multirow{2}{*}{$\begin{pmatrix}5455&-64 \\ -64 & 5480 \end{pmatrix}$}
               & \multirow{2}{*}{$\begin{bmatrix}5403 \\5533 \end{bmatrix}$}  & \multirow{2}{*}{$\begin{bmatrix}(-0.772, -0.635)\\(0.635, -0.772) \end{bmatrix}$}\\
            &  $|\{cc\}^6_0 (\bar c \bar n)^{\bar 6}_0\rangle_0$    \\
 $1^{+}$  &  $ |\{cc\}^{\bar 3}_1 (\bar c \bar n)^3_0\rangle_1  $    & \multirow{3}{*}{$\begin{pmatrix}5425&5 &-37 \\5&5473&6 \\ -37& 6 & 5457  \end{pmatrix}$}
               & \multirow{3}{*}{$\begin{bmatrix}5400 \\5473 \\ 5482 \end{bmatrix}$}  & \multirow{3}{*}{$\begin{bmatrix}(-0.831, 0.105, -0.547)\\(0.231, 0.958, -0.168) \\   (0.506, -0.266, -0.821) \end{bmatrix}$}\\
                 &  $|\{cc\}^{\bar 3}_1 (\bar c \bar n)^3_1\rangle_1$    \\
            &  $|\{cc\}^{6}_0 (\bar c \bar n)^{\bar 6}_1\rangle_1$    \\
 $2^{+}$  &   $|\{cc\}^{\bar 3}_1 (\bar c \bar n)^3_1\rangle_2$    & 5506  &  5506  &  1 \\\hline

 $0^{+}$  & $|\{cc\}^{\bar 3}_1 (\bar c \bar s)^3_1 \rangle_0$    & \multirow{2}{*}{$\begin{pmatrix}5535&62 \\ 62 & 5541 \end{pmatrix}$}
               & \multirow{2}{*}{$\begin{bmatrix}5476 \\5599 \end{bmatrix}$}  & \multirow{2}{*}{$\begin{bmatrix}(-0.723, 0.690)\\(-0.690, -0.723) \end{bmatrix}$}\\
            &  $|\{cc\}^6_0 (\bar c \bar s)^{\bar 6}_0\rangle_0$    \\
 $1^{+}$  &  $ |\{cc\}^{\bar 3}_1 (\bar c \bar s)^3_0\rangle_1  $    & \multirow{3}{*}{$\begin{pmatrix}5512&-4 &-36 \\-4&5552&-4 \\ -36& -4 & 5523  \end{pmatrix}$}
               & \multirow{3}{*}{$\begin{bmatrix}5481 \\5552 \\ 5554 \end{bmatrix}$}  & \multirow{3}{*}{$\begin{bmatrix}(-0.752, -0.078, -0.654)\\(0.346, -0.892, -0.291) \\   (-0.561, -0.445, 0.698) \end{bmatrix}$}\\
                 &  $|\{cc\}^{\bar 3}_1 (\bar c \bar s)^3_1\rangle_1$    \\
            &  $|\{cc\}^{6}_0 (\bar c \bar s)^{\bar 6}_1\rangle_1$    \\
 $2^{+}$  &   $|\{cc\}^{\bar 3}_1 (\bar c \bar s)^3_1\rangle_2$    & 5584  &  5584  &  1 \\\hline

 $0^{+}$  & $|\{cc\}^{\bar 3}_1 (\bar b \bar n)^3_1 \rangle_0$    & \multirow{2}{*}{$\begin{pmatrix}8713&52 \\ 52 & 8707 \end{pmatrix}$}
               & \multirow{2}{*}{$\begin{bmatrix}8658 \\8761 \end{bmatrix}$}  & \multirow{2}{*}{$\begin{bmatrix}(-0.687, 0.727)\\(0.727, 0.687) \end{bmatrix}$}\\
            &  $|\{cc\}^6_0 (\bar b \bar n)^{\bar 6}_0\rangle_0$    \\
 $1^{+}$  &  $ |\{cc\}^{\bar 3}_1 (\bar b \bar n)^3_0\rangle_1  $    & \multirow{3}{*}{$\begin{pmatrix}8716&11 &-30 \\11&8727&20 \\ -30& 20 & 8697  \end{pmatrix}$}
               & \multirow{3}{*}{$\begin{bmatrix}8667 \\8734 \\ 8740 \end{bmatrix}$}  & \multirow{3}{*}{$\begin{bmatrix}(0.537, -0.354, 0.765)\\(-0.651, -0.751, 0.109) \\   (-0.536, 0.557, 0.634) \end{bmatrix}$}\\
                 &  $|\{cc\}^{\bar 3}_1 (\bar b \bar n)^3_1\rangle_1$    \\
            &  $|\{cc\}^{6}_0 (\bar b \bar n)^{\bar 6}_1\rangle_1$    \\
 $2^{+}$  &   $|\{cc\}^{\bar 3}_1 (\bar b \bar n)^3_1\rangle_2$    & 8755  &  8755  &  1 \\\hline

 $0^{+}$  & $|\{cc\}^{\bar 3}_1 (\bar b \bar s)^3_1 \rangle_0$    & \multirow{2}{*}{$\begin{pmatrix}8802&-47 \\ -47 & 8795 \end{pmatrix}$}
               & \multirow{2}{*}{$\begin{bmatrix}8751 \\8845 \end{bmatrix}$}  & \multirow{2}{*}{$\begin{bmatrix}(0.681, 0.732)\\(0.732, -0.681) \end{bmatrix}$}\\
            &  $|\{cc\}^6_0 (\bar b \bar s)^{\bar 6}_0\rangle_0$    \\
 $1^{+}$  &  $ |\{cc\}^{\bar 3}_1 (\bar b \bar s)^3_0\rangle_1  $    & \multirow{3}{*}{$\begin{pmatrix}8805&-9 &27 \\-9&8815&16 \\ 27& 16 & 8786  \end{pmatrix}$}
               & \multirow{3}{*}{$\begin{bmatrix}8761 \\8820 \\ 8825 \end{bmatrix}$}  & \multirow{3}{*}{$\begin{bmatrix}(0.541, 0.322, -0.777)\\(-0.509, 0.861, 0.002) \\   (0.669, 0.395, 0.630) \end{bmatrix}$}\\
                 &  $|\{cc\}^{\bar 3}_1 (\bar b \bar s)^3_1\rangle_1$    \\
            &  $|\{cc\}^{6}_0 (\bar b \bar s)^{\bar 6}_1\rangle_1$    \\
 $2^{+}$  &   $|\{cc\}^{\bar 3}_1 (\bar b \bar s)^3_1\rangle_2$    & 8840  &  8840  &  1 \\

\hline\hline
\end{tabular*}
\end{center}
\end{table*}

By replacing the $\bar c$ in the $cc\bar c \bar q$ tetraquarks with $\bar b$, one can easily obtain the mass spectra of the $cc\bar b \bar q$ system. The masses of the flavor exotic $cc\bar b \bar q$ states vary from 8658 to 8845 MeV, and the mass splittings are about 100 MeV. 
It can be seen that the spectra for the $cc\bar b \bar q$ states show similar patterns to those of $cc\bar c \bar q$ tetraquarks. Also, the mixing effects between different configurations are significant owing to the nondiagonal elements induced by spin-spin interactions. Finally, our results are quite different from those estimated by the color-magnetic interaction model and QCD sum rule, where significantly lower masses are predicted~\cite{Chen:2016ont,Jiang:2017tdc}.

In the present study, the predicted masses for the $cc\bar c \bar q$ and $cc\bar b \bar q$ systems are much higher than the relevant thresholds, and no stable state exists. Hence, the $cc\bar c \bar q$ and $cc\bar b \bar q$ tetraquarks may easily decay into two mesons through the fall-apart mechanism. The possible decay channels of $cc\bar c \bar q$ and $cc\bar b \bar q$ systems via the fall-apart mechanism are listed in Table~\ref{decay1}. We show that the decay channels with two $S-$wave mesons should make dominant contributions; future experiments can search for $cc\bar c \bar q$ and $cc\bar b \bar q$ tetraquarks in these final states. 

\begin{table*}[htb]
\begin{center}
\caption{\label{decay1} The possible decay channels of the $cc\bar c \bar q$ and $cc\bar b \bar q$ systems via the fall-apart mechanism. }
\begin{tabular*}{18cm}{@{\extracolsep{\fill}}p{0.9cm}<{\centering}p{0.9cm}<{\centering}p{2cm}<{\centering}p{10cm}<{\centering}}
\hline\hline
 System  & $J^{P}$ & $S-$wave  &  $P-$wave \\\hline
 $cc\bar c \bar n$  & $0^{+}$ & $\eta_c D$, $J/\psi D^*$  & $\eta_c D_1^{(\prime)}$, $J/\psi D_0^*$, $J/\psi D_1^{(\prime)}$, $J/\psi D_2^*$, $h_c D^{(*)}$, $\chi_{c0} D^*$, $\chi_{c1} D^{(*)}$, $\chi_{c2} D^*$  \\
  & $1^{+}$ & $\eta_c D^*$, $J/\psi D^{(*)}$  &  $\eta_c D_0^*$, $\eta_c D_1^{(\prime)}$, $\eta_c D_2^*$, $J/\psi D_0^*$, $J/\psi D_1^{(\prime)}$, $J/\psi D_2^*$, $h_c D^{(*)}$, $\chi_{c0,1,2} D^{(*)}$  \\
  & $2^{+}$ & $J/\psi D^*$  &  $\eta_c D_1^{(\prime)}$,  $\eta_c D_2^*$, $J/\psi D_0^*$, $J/\psi D_1^{(\prime)}$, $J/\psi D_2^*$, $h_c D^{(*)}$, $\chi_{c0,1,2} D^{(*)}$ \\\hline
 $cc\bar c \bar s$  & $0^{+}$ & $\eta_c D_s$, $J/\psi D^*_s$  & $\eta_c D_{s1}^{(\prime)}$, $J/\psi D_{s0}^*$, $J/\psi D_{s1}^{(\prime)}$, $J/\psi D_{s2}^*$, $h_c D_s^{(*)}$, $\chi_{c0} D_s^*$, $\chi_{c1} D_s^{(*)}$, $\chi_{c2} D_s^*$  \\
  & $1^{+}$ & $\eta_c D_s^*$, $J/\psi D_s^{(*)}$  &  $\eta_c D_{s0}^*$, $\eta_c D_{s1}^{(\prime)}$, $\eta_c D_{s2}^*$, $J/\psi D_{s0}^*$, $J/\psi D_{s1}^{(\prime)}$, $J/\psi D_{s2}^*$, $h_c D_s^{(*)}$, $\chi_{c0,1,2} D_s^{(*)}$  \\
  & $2^{+}$ & $J/\psi D_s^*$  &  $\eta_c D_{s1}^{(\prime)}$, $\eta_c D_{s2}^*$, $J/\psi D_{s0}^*$, $J/\psi D_{s1}^{(\prime)}$, $J/\psi D_{s2}^*$, $h_c D_s^{(*)}$, $\chi_{c0,1,2} D_s^{(*)}$ \\\hline
 $cc\bar b \bar n$  & $0^{+}$ & $B_c D$, $B_c^* D^*$  & $B_c D_1^{(\prime)}$, $B_c^* D_0^*$, $B_c^* D_1^{(\prime)}$, $B_c^* D_2^*$,  $B_{c0}^* D^*$, $B_{c1}^{(\prime)} D^{(*)}$,  $B_{c2}^* D^*$  \\
  & $1^{+}$ & $B_c D^*$, $B_c^* D^{(*)}$  &  $B_c D_0^*$, $B_c D_1^{(\prime)}$, $B_c D_2^*$, $B_c^* D_0^*$, $B_c^* D_1^{(\prime)}$, $B_c^* D_2^*$, $B_{c0}^* D^{(*)}$, $B_{c1}^{(\prime)} D^{(*)}$, $B_{c2}^* D^{(*)}$  \\
  & $2^{+}$ & $B_c^* D^*$  & $B_c D_1^{(\prime)}$, $B_c D_2^*$, $B_c^* D_0^*$, $B_c^* D_1^{(\prime)}$, $B_c^* D_2^*$, $B_{c0}^* D^{(*)}$, $B_{c1}^{(\prime)} D^{(*)}$, $B_{c2}^* D^{(*)}$  \\\hline
 $cc\bar b \bar s$  & $0^{+}$ & $B_c D_s$, $B_c^* D_s^*$  & $B_c D_{s1}^{(\prime)}$, $B_c^* D_{s0}^*$, $B_c^* D_{s1}^{(\prime)}$, $B_c^* D_{s2}^*$,  $B_{c0}^* D_s^*$, $B_{c1}^{(\prime)} D_s^{(*)}$,  $B_{c2}^* D_s^*$  \\
  & $1^{+}$ & $B_c D_s^*$, $B_c^* D_s^{(*)}$  &  $B_c D_{s0}^*$, $B_c D_{s1}^{(\prime)}$, $B_c D_{s2}^*$, $B_c^* D_{s0}^*$, $B_c^* D_{s1}^{(\prime)}$, $B_c^* D_{s2}^*$, $B_{c0}^* D_s^{(*)}$, $B_{c1}^{(\prime)} D_s^{(*)}$, $B_{c2}^* D_s^{(*)}$  \\
  & $2^{+}$ & $B_c^* D_s^*$  & $B_c D_{s1}^{(\prime)}$, $B_c D_{s2}^*$, $B_c^* D_{s0}^*$, $B_c^* D_{s1}^{(\prime)}$, $B_c^* D_{s2}^*$, $B_{c0}^* D_s^{(*)}$, $B_{c1}^{(\prime)} D_s^{(*)}$, $B_{c2}^* D_s^{(*)}$  \\
\hline\hline
\end{tabular*}
\end{center}
\end{table*}

\subsection{The $bb\bar c \bar q$ and $bb\bar b \bar q$ systems}

The masses of $bb\bar c \bar q$ and $bb\bar b \bar q$ systems are presented in Table~\ref{mass2}. In the $bb\bar c \bar q$ system, the tetraquarks are flavor exotic and cannot mix with the conventional mesons. The predicted mass spectra lie around 12 GeV, and the mass splittings are 140 and 117 MeV for the $bb\bar c \bar n$ and $bb\bar c \bar s$ tetraquark states, respectively. In the hidden bottom systems, the masses of $bb\bar b \bar q$ tetraquarks vary from 15107 to 15232 MeV, and the mass splittings are relatively small. Since these masses are much greater than those of the conventional bottom or bottom-strange mesons, it is not necessary to consider the mixing effects between them. The possible decay modes for $bb\bar c \bar q$ and $bb\bar b \bar q$ systems via the fall-apart mechanism are presented in Table~\ref{decay2}. As they have a large phase space, these tetraquarks may easily decay into two $S-$wave mesons, which can be searched for by future experiments.

\begin{table*}[htp]
\begin{center}
\caption{\label{mass2} Predicted mass spectra for the $bb \bar c \bar q$ and $bb \bar b \bar q$ tetraquarks.}
\begin{tabular*}{18cm}{@{\extracolsep{\fill}}p{1cm}<{\centering}p{1.7cm}<{\centering}p{4.5cm}<{\centering}p{1.8cm}<{\centering}p{4.5cm}<{\centering}}
\hline\hline
 $J^{P}$  & Configuration                                             & $\langle H\rangle$ (MeV) & Mass (MeV)  & Eigenvector\\\hline

 $0^{+}$  & $|\{bb\}^{\bar 3}_1 (\bar c \bar n)^3_1 \rangle_0$    & \multirow{2}{*}{$\begin{pmatrix}11925&-28 \\ -28 & 12004 \end{pmatrix}$}
               & \multirow{2}{*}{$\begin{bmatrix}11916 \\12013 \end{bmatrix}$}  & \multirow{2}{*}{$\begin{bmatrix}(-0.954, -0.299)\\(0.299, -0.954) \end{bmatrix}$}\\
            &  $|\{bb\}^6_0 (\bar c \bar n)^{\bar 6}_0\rangle_0$    \\
 $1^{+}$  &  $ |\{bb\}^{\bar 3}_1 (\bar c \bar n)^3_0\rangle_1  $    & \multirow{3}{*}{$\begin{pmatrix}11875&-1 &-16 \\-1&11933&1 \\ -16& 1 & 11982  \end{pmatrix}$}
               & \multirow{3}{*}{$\begin{bmatrix}11873 \\11933 \\ 11984 \end{bmatrix}$}  & \multirow{3}{*}{$\begin{bmatrix}(-0.989, -0.008, -0.149)\\(0.004, -1.000, 0.028) \\   (-0.150, 0.027, 0.988) \end{bmatrix}$}\\
                 &  $|\{bb\}^{\bar 3}_1 (\bar c \bar n)^3_1\rangle_1$    \\
            &  $|\{bb\}^{6}_0 (\bar c \bar n)^{\bar 6}_1\rangle_1$    \\
 $2^{+}$  &   $|\{bb\}^{\bar 3}_1 (\bar c \bar n)^3_1\rangle_2$    & 11948  &  11948  &  1 \\\hline

 $0^{+}$  & $|\{bb\}^{\bar 3}_1 (\bar c \bar s)^3_1 \rangle_0$    & \multirow{2}{*}{$\begin{pmatrix}11995&27 \\ 27 & 12056 \end{pmatrix}$}
               & \multirow{2}{*}{$\begin{bmatrix}11985 \\12067 \end{bmatrix}$}  & \multirow{2}{*}{$\begin{bmatrix}(-0.935, 0.355)\\(-0.355, -0.935) \end{bmatrix}$}\\
            &  $|\{bb\}^6_0 (\bar c \bar s)^{\bar 6}_0\rangle_0$    \\
 $1^{+}$  &  $ |\{bb\}^{\bar 3}_1 (\bar c \bar s)^3_0\rangle_1  $    & \multirow{3}{*}{$\begin{pmatrix}11953&0 &16 \\0&12003&1 \\ 16& 1 & 12037  \end{pmatrix}$}
               & \multirow{3}{*}{$\begin{bmatrix}11950 \\12003 \\ 12040 \end{bmatrix}$}  & \multirow{3}{*}{$\begin{bmatrix}(-0.983, 0.002, 0.183)\\(0.004, -0.999, 0.036) \\   (0.183, 0.036, 0.983) \end{bmatrix}$}\\
                 &  $|\{bb\}^{\bar 3}_1 (\bar c \bar s)^3_1\rangle_1$    \\
            &  $|\{bb\}^{6}_0 (\bar c \bar s)^{\bar 6}_1\rangle_1$    \\
 $2^{+}$  &   $|\{bb\}^{\bar 3}_1 (\bar c \bar s)^3_1\rangle_2$    & 12018  &  12018  &  1 \\\hline

 $0^{+}$  & $|\{bb\}^{\bar 3}_1 (\bar b \bar n)^3_1 \rangle_0$    & \multirow{2}{*}{$\begin{pmatrix}15134&24 \\ 24 & 15133 \end{pmatrix}$}
               & \multirow{2}{*}{$\begin{bmatrix}15109 \\15158 \end{bmatrix}$}  & \multirow{2}{*}{$\begin{bmatrix}(-0.694, 0.720)\\(0.720, 0.694) \end{bmatrix}$}\\
            &  $|\{bb\}^6_0 (\bar b \bar n)^{\bar 6}_0\rangle_0$    \\
 $1^{+}$  &  $ |\{bb\}^{\bar 3}_1 (\bar b \bar n)^3_0\rangle_1  $    & \multirow{3}{*}{$\begin{pmatrix}15123&3 &14 \\3&15141&-4 \\ 14& -4 & 15122  \end{pmatrix}$}
               & \multirow{3}{*}{$\begin{bmatrix}15107 \\15137 \\ 15142 \end{bmatrix}$}  & \multirow{3}{*}{$\begin{bmatrix}(0.693, -0.155, -0.705)\\(-0.721, -0.151, -0.676) \\   (0.002, -0.976, 0.217) \end{bmatrix}$}\\
                 &  $|\{bb\}^{\bar 3}_1 (\bar b \bar n)^3_1\rangle_1$    \\
            &  $|\{bb\}^{6}_0 (\bar b \bar n)^{\bar 6}_1\rangle_1$    \\
 $2^{+}$  &   $|\{bb\}^{\bar 3}_1 (\bar b \bar n)^3_1\rangle_2$    & 15154  &  15154  &  1 \\\hline

 $0^{+}$  & $|\{bb\}^{\bar 3}_1 (\bar b \bar s)^3_1 \rangle_0$    & \multirow{2}{*}{$\begin{pmatrix}15213&23 \\ 23 & 15204 \end{pmatrix}$}
               & \multirow{2}{*}{$\begin{bmatrix}15185 \\15232 \end{bmatrix}$}  & \multirow{2}{*}{$\begin{bmatrix}(-0.641, 0.767)\\(0.767, 0.641) \end{bmatrix}$}\\
            &  $|\{bb\}^6_0 (\bar b \bar s)^{\bar 6}_0\rangle_0$    \\
 $1^{+}$  &  $ |\{bb\}^{\bar 3}_1 (\bar b \bar s)^3_0\rangle_1  $    & \multirow{3}{*}{$\begin{pmatrix}15203&-2 &-14 \\-2&15219&-3 \\ -14& -3 & 15195  \end{pmatrix}$}
               & \multirow{3}{*}{$\begin{bmatrix}15184 \\15213 \\ 15219 \end{bmatrix}$}  & \multirow{3}{*}{$\begin{bmatrix}(-0.607, -0.113, -0.787)\\(-0.791, -0.009, 0.611) \\   (0.076, -0.994, 0.084) \end{bmatrix}$}\\
                 &  $|\{bb\}^{\bar 3}_1 (\bar b \bar s)^3_1\rangle_1$    \\
            &  $|\{bb\}^{6}_0 (\bar b \bar s)^{\bar 6}_1\rangle_1$    \\
 $2^{+}$  &   $|\{bb\}^{\bar 3}_1 (\bar b \bar s)^3_1\rangle_2$    & 15231  &  15231  &  1 \\

\hline\hline
\end{tabular*}
\end{center}
\end{table*}

\begin{table*}[htb]
\begin{center}
\caption{\label{decay2} The possible decay channels of the $bb\bar c \bar q$ and $bb\bar b \bar q$ systems via the fall-apart mechanism. }
\begin{tabular*}{18cm}{@{\extracolsep{\fill}}p{0.9cm}<{\centering}p{0.9cm}<{\centering}p{2cm}<{\centering}p{10cm}<{\centering}}
\hline\hline
 System  & $J^{P}$ & $S-$wave  &  $P-$wave \\\hline
$bb\bar c \bar n$ & $0^{+}$ & $\bar B_c \bar B$, $\bar B_c^* \bar B^*$ & $\bar B_c \bar B_1^{(\prime)}$, $\bar B_c^* \bar B_0^*$, $\bar B_c^* \bar B_1^{(\prime)}$, $\bar B_c^* \bar B_2^*$, $\bar B_{c0}^* \bar B^*$, $\bar B_{c1}^{(\prime)} \bar B^{(*)}$, $\bar B_{c2}^* \bar B^*$ \\
& $1^{+}$ & $\bar B_c \bar B^*$, $\bar B_c^* \bar B^{(*)}$ & $\bar B_c \bar B_0^*$, $\bar B_c \bar B_1^{(\prime)}$, $\bar B_c \bar B_2^*$, $\bar B_c^* \bar B_0^*$, $\bar B_c^* \bar B_1^{(\prime)}$, $\bar B_c^* \bar B_2^*$, $\bar B_{c0}^* \bar B^{(*)}$, $\bar B_{c1}^{(\prime)} \bar B^{(*)}$, $\bar B_{c2}^* \bar B^{(*)}$ \\
& $2^{+}$ & $\bar B_c^* \bar B^*$ & $\bar B_c \bar B_1^{(\prime)}$, $\bar B_c \bar B_2^*$, $\bar B_c^* \bar B_0^*$, $\bar B_c^* \bar B_1^{(\prime)}$, $\bar B_c^* \bar B_2^*$, $\bar B_{c0}^* \bar B^{(*)}$, $\bar B_{c1}^{(\prime)} \bar B^{(*)}$, $\bar B_{c2}^* \bar B^{(*)}$ \\\hline
$bb\bar c \bar s$ & $0^{+}$ & $\bar B_c \bar B_s$, $\bar B_c^* \bar B_s^*$ & $\bar B_c \bar B_{s1}^{(\prime)}$, $\bar B_c^* \bar B_{s0}^*$, $\bar B_c^* \bar B_{s1}^{(\prime)}$, $\bar B_c^* \bar B_{s2}^*$, $\bar B_{c0}^* \bar B_s^*$, $\bar B_{c1}^{(\prime)} \bar B_s^{(*)}$, $\bar B_{c2}^* \bar B_s^*$ \\
& $1^{+}$ & $\bar B_c \bar B_s^*$, $\bar B_c^* \bar B_s^{(*)}$ & $\bar B_c \bar B_{s0}^*$, $\bar B_c \bar B_{s1}^{(\prime)}$, $\bar B_c \bar B_{s2}^*$, $\bar B_c^* \bar B_{s0}^*$, $\bar B_c^* \bar B_{s1}^{(\prime)}$, $\bar B_c^* \bar B_{s2}^*$, $\bar B_{c0}^* \bar B_s^{(*)}$, $\bar B_{c1}^{(\prime)} \bar B_s^{(*)}$, $\bar B_{c2}^* \bar B_s^{(*)}$ \\
& $2^{+}$ & $\bar B_c^* \bar B_s^*$ & $\bar B_c \bar B_{s1}^{(\prime)}$, $\bar B_c \bar B_{s2}^*$, $\bar B_c^* \bar B_{s0}^*$, $\bar B_c^* \bar B_{s1}^{(\prime)}$, $\bar B_c^* \bar B_{s2}^*$, $\bar B_{c0}^* \bar B_s^{(*)}$, $\bar B_{c1}^{(\prime)} \bar B_s^{(*)}$, $\bar B_{c2}^* \bar B_s^{(*)}$ \\\hline
$bb\bar b \bar n$ & $0^{+}$ & $\eta_b \bar B$, $\Upsilon \bar B^*$ & $\eta_b \bar B_1^{(\prime)}$, $\Upsilon \bar B_0^*$, $\Upsilon \bar B_1^{(\prime)}$, $\Upsilon \bar B_2^*$, $h_b \bar B^{(*)}$, $\chi_{b0} \bar B^*$, $\chi_{b1} \bar B^{(*)}$, $\chi_{b2} \bar B^*$ \\
& $1^{+}$ & $\eta_b \bar B^*$, $\Upsilon \bar B^{(*)}$ & $\eta_b \bar B_0^*$, $\eta_b \bar B_1^{(\prime)}$, $\eta_b \bar B_2^*$, $\Upsilon \bar B_0^*$, $\Upsilon \bar B_1^{(\prime)}$, $\Upsilon \bar B_2^*$, $h_b \bar B^{(*)}$, $\chi_{b0,1,2} \bar B^{(*)}$ \\
& $2^{+}$ & $\Upsilon \bar B^*$ & $\eta_b \bar B_1^{(\prime)}$, $\eta_b \bar B_2^*$, $\Upsilon \bar B_0^*$, $\Upsilon \bar B_1^{(\prime)}$, $\Upsilon \bar B_2^*$, $h_b \bar B^{(*)}$, $\chi_{b0,1,2} \bar B^{(*)}$ \\\hline
$bb\bar b \bar s$ & $0^{+}$ & $\eta_b \bar B_s$, $\Upsilon \bar B^*_s$ & $\eta_b \bar B_{s1}^{(\prime)}$, $\Upsilon \bar B_{s0}^*$, $\Upsilon \bar B_{s1}^{(\prime)}$, $\Upsilon \bar B_{s2}^*$, $h_b \bar B_s^{(*)}$, $\chi_{b0} \bar B_s^*$, $\chi_{b1} \bar B_s^{(*)}$, $\chi_{b2} \bar B_s^*$ \\
& $1^{+}$ & $\eta_b \bar B_s^*$, $\Upsilon \bar B_s^{(*)}$ & $\eta_b \bar B_{s0}^*$, $\eta_b \bar B_{s1}^{(\prime)}$, $\eta_b \bar B_{s2}^*$, $\Upsilon \bar B_{s0}^*$, $\Upsilon \bar B_{s1}^{(\prime)}$, $\Upsilon \bar B_{s2}^*$, $h_b \bar B_s^{(*)}$, $\chi_{b0,1,2} \bar B_s^{(*)}$ \\
& $2^{+}$ & $\Upsilon \bar B_s^*$ & $\eta_b \bar B_{s1}^{(\prime)}$, $\eta_b \bar B_{s2}^*$, $\Upsilon \bar B_{s0}^*$, $\Upsilon \bar B_{s1}^{(\prime)}$, $\Upsilon \bar B_{s2}^*$, $h_b \bar B_s^{(*)}$, $\chi_{b0,1,2} \bar B_s^{(*)}$ \\
\hline\hline
\end{tabular*}
\end{center}
\end{table*}

Our predicted masses for the $bb\bar c \bar q$ and $bb\bar b \bar q$ tetraquarks are about several hundred MeV higher than those predicted by the previous studies within the color-magnetic interaction model and QCD sum rule~\cite{Chen:2016ont,Jiang:2017tdc}. The differences among these studies arise from the different choices of interactions, parameters, or even models. Indeed, only the color-magnetic interactions are considered in Ref.~\cite{Chen:2016ont}, while all the Coulomb potential, confining potential, spin-spin interactions, and the relativistic corrections are involved in the present study. Also, the input parameters are quite different in the color-magnetic interaction model and the extended relativized quark model. Moreover, it is difficult to establish a clear connection between the constituent quark model and QCD sum rule. Further investigations with various theoretical approaches and experimental searches are encouraged in order to disentangle these differences.

As in the $cc\bar c \bar q$ and $cc\bar b \bar q$ systems, all the mass splittings in the $bb\bar c \bar q$ and $bb\bar b \bar q$ systems are small. This is easy to understand from the Hamitonian, where only the spin-spin interactions contribute to the mass splittings. The spin-spin terms are suppressed by the heavy quark masses and may even cancel each other out, which makes the mass splittings for  $cc\bar c \bar q$, $cc\bar b \bar q$, $bb\bar c \bar q$, and $bb\bar b \bar q$ states similar to those of conventional heavy-light mesons.

\subsection{The $cb\bar c \bar q$ and $cb\bar b \bar q$ systems}

The predicted masses of the $cb\bar c \bar q$ and $cb\bar b \bar q$ systems are listed in Table~\ref{mass3}. This shows that all the $cb\bar c \bar q$ and $cb\bar b \bar q$ states lie above the coresponding meson-meson thresholds. Unlike the $cc\bar c \bar q$, $cc\bar b \bar q$ , $bb\bar c \bar q$, and $bb\bar b \bar q$ systems, several characteristics appear for the $cb\bar c \bar q$ and $cb\bar b \bar q$ tetraquark states. All of these $cb\bar c \bar q$ and $cb\bar b \bar q$ states are hidden charm or bottom tetraquarks, and no flavor exotic state exists. In principle, they can be regarded as the excited $D$, $D_s$, $B$, and $B_s$ mesons with an extra $c\bar c$ or $b \bar b$ quark pair. Certainly, the much higher masses prevent mixing between these tetraquarks and conventional mesons. Thus, without the constraint from the Pauli exclusion principle, the number of allowed states is doubled, as are the meson-meson thresholds. The possible fall-apart decay modes are presented in Table~\ref{decay3}, and the channels with two $S-$wave mesons should play essential roles. These two-body decay modes may provide helpful information for future experimental searches.

\begin{table*}[htp]
\begin{center}
\caption{\label{mass3} Predicted mass spectra for the $cb \bar c \bar q$ and $cb \bar b \bar q$ tetraquarks.}
\footnotesize
\begin{tabular*}{18cm}{@{\extracolsep{\fill}}p{0.15cm}<{\centering}p{0.7cm}<{\centering}p{6.4cm}<{\centering}p{1.5cm}<{\centering}p{6.4cm}<{\centering}}
\hline\hline
 $J^{P}$  & Configuration                                             & $\langle H\rangle$ (MeV) & Mass (MeV)  & Eigenvector\\\hline
 $0^{+}$  & $|(cb)^{\bar 3}_0 (\bar c \bar n)^3_0 \rangle_0$    & \multirow{4}{*}{$\begin{pmatrix}8660&-3 &73 & -48 \\-3&8719&43&-86 \\ 73& 43 & 8764 & 10 \\ -48 & -86 & 10 & 8666  \end{pmatrix}$}
               & \multirow{4}{*}{$\begin{bmatrix}8560 \\8663 \\ 8755 \\ 8831 \end{bmatrix}$}  & \multirow{4}{*}{$\begin{bmatrix}(0.549, 0.440, -0.323, 0.633)\\(0.716, -0.448, -0.284, -0.454) \\            (0.200, -0.586, 0.580, 0.530) \\ (0.382, 0.512, 0.692, -0.335) \end{bmatrix}$}\\
                 &  $|(cb)^{\bar 3}_1 (\bar c \bar n)^3_1\rangle_0$    \\
            &  $|(cb)^6_0 (\bar c \bar n)^{\bar 6}_0\rangle_0$    \\
            &  $|(cb)^6_1 (\bar c \bar n)^{\bar 6}_1\rangle_0$    \\
 $1^{+}$  &  $|(cb)^{\bar 3}_0 (\bar c \bar n)^3_1\rangle_1$   & \multirow{6}{*}{$\begin{pmatrix}8725&-2 &6 & 75 &  25 & 2 \\-2&8678&3&25 &75  &10 \\ 6& 3 & 8731 & -3 & -10& -81 \\ 75 & 25 & -3 & 8741 & -6 & -10 \\ 25&75 &-10 &  -6 & 8760 & 0 \\ 2 &10 & -81 &  -10 & 0 & 8703 \\ \end{pmatrix}$}
               &\multirow{6}{*}{$\begin{bmatrix}8607 \\8643 \\ 8674 \\ 8782 \\ 8805 \\8826 \end{bmatrix}$}  & \multirow{6}{*}{$\begin{bmatrix}(0.334, 0.609, -0.317, -0.353, -0.386, -0.380)\\(0.156, 0.382, 0.559, -0.149, -0.237, 0.662) \\   (-0.660, 0.488, -0.038, 0.529, -0.207, -0.049) \\ (-0.332, 0.293, 0.392, -0.458, 0.584, -0.317) \\    (-0.239, 0.143, -0.654, -0.312, 0.290, 0.560) \\ (-0.511, -0.373, 0.063, -0.516, -0.572, -0.039) \\ \end{bmatrix}$}\\
               & $|(cb)^{\bar 3}_1 (\bar c \bar n)^3_0\rangle_1 $  \\
                 &  $|(cb)^{\bar 3}_1 (\bar c \bar n)^3_1\rangle_1$    \\
            &  $|(cb)^6_0 (\bar c \bar n)^{\bar 6}_1\rangle_1$   \\
           & $|(cb)^6_1 (\bar c \bar n)^{\bar 6}_0\rangle_1$    \\
           &  $|(cb)^6_1 (\bar c \bar n)^{\bar 6}_1\rangle_1$    \\
 $2^{+}$  &  $|(cb)^{\bar 3}_1 (\bar c \bar n)^3_1\rangle_2$    & \multirow{2}{*}{$\begin{pmatrix}8754&-72\\ -72 & 8769 \\ \end{pmatrix}$}
               & \multirow{2}{*}{$\begin{bmatrix}8690 \\8834 \end{bmatrix}$}  & \multirow{2}{*}{$\begin{bmatrix}(-0.744, -0.668)\\(0.668, -0.744) \\\end{bmatrix}$}\\
            &  $|(cb)^6_1 (\bar c \bar n)^{\bar 6}_1\rangle_2$   \\\hline

 $0^{+}$  & $|(cb)^{\bar 3}_0 (\bar c \bar s)^3_0 \rangle_0$    & \multirow{4}{*}{$\begin{pmatrix}8743&-2 &61 & 46 \\-2&8794&42&72 \\ 61& 42 & 8823 & -8 \\ 46 & 72 & -8 & 8733  \end{pmatrix}$}
               & \multirow{4}{*}{$\begin{bmatrix}8646 \\8748 \\ 8811 \\ 8887 \end{bmatrix}$}  & \multirow{4}{*}{$\begin{bmatrix}(0.530, 0.421, -0.315, -0.666)\\(-0.722, 0.406, 0.311, -0.466) \\            (-0.229, 0.602, -0.596, 0.480) \\ (0.381, 0.544, 0.670, 0.331) \end{bmatrix}$}\\
                 &  $|(cb)^{\bar 3}_1 (\bar c \bar s)^3_1\rangle_0$    \\
            &  $|(cb)^6_0 (\bar c \bar s)^{\bar 6}_0\rangle_0$    \\
            &  $|(cb)^6_1 (\bar c \bar s)^{\bar 6}_1\rangle_0$    \\
 $1^{+}$  &  $|(cb)^{\bar 3}_0 (\bar c \bar s)^3_1\rangle_1$   & \multirow{6}{*}{$\begin{pmatrix}8799&1 &6 & 62 &  24 & -1 \\1&8761&-2&-25 &-62  &10 \\ 6& -2 & 8806 & -2 & -9& 67 \\ 62 & -25 & -2 & 8804 & -5 & 9 \\ 24&-62 &-9 &  -5 & 8819 & 0 \\ -1 &10 &67 &  9 & 0 & 8767 \\ \end{pmatrix}$}
               & \multirow{6}{*}{$\begin{bmatrix}8694 \\8725 \\ 8757 \\ 8836 \\ 8862 \\8882 \end{bmatrix}$}  & \multirow{6}{*}{$\begin{bmatrix}(-0.335, 0.571, 0.316, 0.373, 0.385, -0.417)\\(0.197, -0.367, 0.512, -0.192, -0.255, -0.680) \\   (0.625, 0.517, 0.018, -0.530, 0.246, -0.027) \\ (0.352, 0.325, -0.373, 0.457, -0.595, -0.260) \\    (0.221, 0.139, 0.704, 0.269, -0.265, 0.542) \\ (-0.534, 0.383, 0.057, -0.512, -0.550, 0.032) \\ \end{bmatrix}$}\\
               & $|(cb)^{\bar 3}_1 (\bar c \bar s)^3_0\rangle_1 $  \\
                 &  $|(cb)^{\bar 3}_1 (\bar c \bar s)^3_1\rangle_1$    \\
            &  $|(cb)^6_0 (\bar c \bar s)^{\bar 6}_1\rangle_1$   \\
           & $|(cb)^6_1 (\bar c \bar s)^{\bar 6}_0\rangle_1$    \\
           &  $|(cb)^6_1 (\bar c \bar s)^{\bar 6}_1\rangle_1$    \\
 $2^{+}$  &  $|(cb)^{\bar 3}_1 (\bar c \bar s)^3_1\rangle_2$    &\multirow{2}{*}{$\begin{pmatrix}8828&-60\\ -60 & 8831 \\ \end{pmatrix}$}
               & \multirow{2}{*}{$\begin{bmatrix}8770 \\8889 \end{bmatrix}$}  & \multirow{2}{*}{$\begin{bmatrix}(-0.714, -0.700)\\(0.700, -0.714) \\\end{bmatrix}$}\\
            &  $|(cb)^6_1 (\bar c \bar s)^{\bar 6}_1\rangle_2$   \\\hline

 $0^{+}$  & $|(cb)^{\bar 3}_0 (\bar b \bar n)^3_0 \rangle_0$    & \multirow{4}{*}{$\begin{pmatrix}11926&5 &-112 & -39 \\5&11949&36&129 \\ -112& 36 & 11927 & -14 \\ -39 & 129 & -14 & 11854  \end{pmatrix}$}
               & \multirow{4}{*}{$\begin{bmatrix}11735 \\11843 \\ 12010 \\ 12067 \end{bmatrix}$}  & \multirow{4}{*}{$\begin{bmatrix}(-0.368, 0.493, -0.362, -0.701)\\(0.603, 0.286, 0.609, -0.429) \\            (0.510, 0.579, -0.498, 0.396) \\ (-0.491, 0.583, 0.501, 0.410) \end{bmatrix}$}\\
                 &  $|(cb)^{\bar 3}_1 (\bar b \bar n)^3_1\rangle_0$    \\
            &  $|(cb)^6_0 (\bar b \bar n)^{\bar 6}_0\rangle_0$    \\
            &  $|(cb)^6_1 (\bar b \bar n)^{\bar 6}_1\rangle_0$    \\
 $1^{+}$  &  $|(cb)^{\bar 3}_0 (\bar b \bar n)^3_1\rangle_1$   & \multirow{6}{*}{$\begin{pmatrix}11951&-3 &-5 & -112 &  -21 & 12 \\-3&11944&-7&-21 &-114  &9 \\ -5& -7 & 11959 & -12 & -8& 121 \\ -112 & -21 & -12 & 11917 & -8 & 9 \\ -21&-114 &-8 &  -8 & 11922 & 13 \\ 12 &9 & 121 &  9 & 13 & 11884 \\ \end{pmatrix}$}
               & \multirow{6}{*}{$\begin{bmatrix}11767 \\11820 \\ 11846 \\ 12031 \\ 12049 \\12064 \end{bmatrix}$}  & \multirow{6}{*}{$\begin{bmatrix}(0.328, 0.338, 0.415, 0.380, 0.382, -0.562)\\(0.317, 0.341, -0.423, 0.372, 0.365, 0.577) \\   (0.465, -0.470, 0.004, 0.542, -0.519, 0.005) \\ (-0.533, 0.498, 0.197, 0.455, -0.452, 0.135) \\    (0.019, -0.225, 0.765, -0.024, 0.208, 0.566) \\ (-0.540, -0.499, -0.159, 0.466, 0.452, -0.111) \\ \end{bmatrix}$}\\
               & $|(cb)^{\bar 3}_1 (\bar b \bar n)^3_0\rangle_1 $  \\
                 &  $|(cb)^{\bar 3}_1 (\bar b \bar n)^3_1\rangle_1$    \\
            &  $|(cb)^6_0 (\bar b \bar n)^{\bar 6}_1\rangle_1$   \\
           & $|(cb)^6_1 (\bar b \bar n)^{\bar 6}_0\rangle_1$    \\
           &  $|(cb)^6_1 (\bar b \bar n)^{\bar 6}_1\rangle_1$    \\
 $2^{+}$  &  $|(cb)^{\bar 3}_1 (\bar b \bar n)^3_1\rangle_2$    & \multirow{2}{*}{$\begin{pmatrix}11979&-107\\ -107 & 11939 \\ \end{pmatrix}$}
               & \multirow{2}{*}{$\begin{bmatrix}11850 \\12068 \end{bmatrix}$}  & \multirow{2}{*}{$\begin{bmatrix}(0.639, 0.769)\\(0.769, -0.639) \\\end{bmatrix}$}\\
            &  $|(cb)^6_1 (\bar b \bar n)^{\bar 6}_1\rangle_2$   \\\hline

 $0^{+}$  & $|(cb)^{\bar 3}_0 (\bar b \bar s)^3_0 \rangle_0$    & \multirow{4}{*}{$\begin{pmatrix}12010&4 &104 & -36 \\4&12033&-34& 119 \\ 104& -34 & 12007 & 12 \\ -36 & 119 & 12 & 11940  \end{pmatrix}$}
               & \multirow{4}{*}{$\begin{bmatrix}11833 \\11931 \\ 12086 \\ 12140 \end{bmatrix}$}  & \multirow{4}{*}{$\begin{bmatrix}(-0.367, 0.487, 0.362, -0.705)\\(0.595, 0.279, -0.616, -0.433) \\            (0.520, 0.580, 0.495, 0.385) \\ (0.490, -0.590, 0.494, -0.409) \end{bmatrix}$}\\
                 &  $|(cb)^{\bar 3}_1 (\bar b \bar s)^3_1\rangle_0$    \\
            &  $|(cb)^6_0 (\bar b \bar s)^{\bar 6}_0\rangle_0$    \\
            &  $|(cb)^6_1 (\bar b \bar s)^{\bar 6}_1\rangle_0$    \\
 $1^{+}$  &  $|(cb)^{\bar 3}_0 (\bar b \bar s)^3_1\rangle_1$   & \multirow{6}{*}{$\begin{pmatrix}12033&2 &-4 & -104 &  -19 & 9 \\2&12029&6&20 &105  &-7 \\ -4& 6 & 12042 & -9 & -7& 112 \\ -104 & 20 & -9 & 11999 & -7 & 7 \\ -20&105 &-7 &  -7 & 12003 & 10 \\ 9 &-7 & 112 &  7 & 10 & 11968 \\ \end{pmatrix}$}
               & \multirow{6}{*}{$\begin{bmatrix}11865 \\11908 \\ 11934 \\ 12106 \\ 12123 \\12137 \end{bmatrix}$}  & \multirow{6}{*}{$\begin{bmatrix}(0.328, -0.335, 0.411, 0.383, 0.382, -0.564)\\(-0.319, 0.331, 0.416, -0.376, -0.370, -0.583) \\   (0.458, 0.465, 0.004, 0.542, -0.530, 0.001) \\ (-0.544, -0.516, 0.123, 0.459, -0.453, 0.083) \\    (-0.017, 0.172, 0.789, 0.008, 0.151, 0.570) \\ (-0.533, 0.516, -0.144, 0.457, 0.458, -0.102) \\ \end{bmatrix}$}\\
               & $|(cb)^{\bar 3}_1 (\bar b \bar s)^3_0\rangle_1 $  \\
                 &  $|(cb)^{\bar 3}_1 (\bar b \bar s)^3_1\rangle_1$    \\
            &  $|(cb)^6_0 (\bar b \bar s)^{\bar 6}_1\rangle_1$   \\
           & $|(cb)^6_1 (\bar b \bar s)^{\bar 6}_0\rangle_1$    \\
           &  $|(cb)^6_1 (\bar b \bar s)^{\bar 6}_1\rangle_1$    \\
 $2^{+}$  &  $|(cb)^{\bar 3}_1 (\bar b \bar s)^3_1\rangle_2$    & \multirow{2}{*}{$\begin{pmatrix}12061&100\\ 100 & 12019 \\ \end{pmatrix}$}
               & \multirow{2}{*}{$\begin{bmatrix}11938 \\12142 \end{bmatrix}$}  & \multirow{2}{*}{$\begin{bmatrix}(-0.631, 0.776)\\(0.776, 0.631) \\\end{bmatrix}$}\\
            &  $|(cb)^6_1 (\bar b \bar s)^{\bar 6}_1\rangle_2$   \\

\hline\hline
\end{tabular*}
\end{center}
\end{table*}

\begin{table*}[htb]
\begin{center}
\small
\caption{\label{decay3} The possible decay channels of the $cb\bar c \bar q$ and $cb\bar b \bar q$ systems via the fall-apart mechanism. }
\begin{tabular*}{18cm}{@{\extracolsep{\fill}}p{0.8cm}<{\centering}p{0.8cm}<{\centering}p{2.1cm}<{\centering}p{10cm}<{\centering}}
\hline\hline
 System  & $J^{P}$ & $S-$wave  &  $P-$wave \\\hline
$cb\bar c \bar n$ & $0^{+}$ & $\eta_c \bar B$, $J/\psi \bar B^*$, $\bar B_c D$, $\bar B_c^* D^*$ & $\eta_c \bar B_1^{(\prime)}$, $J/\psi \bar B_0^*$, $J/\psi \bar B_1^{(\prime)}$, $J/\psi \bar B_2^*$, $h_c \bar B^{(*)}$, $\chi_{c0} \bar B^*$, $\chi_{c1} \bar B^{(*)}$, $\chi_{c2} \bar B^*$, $\bar B_c D_1^{(\prime)}$, $\bar B_c^* D_0^*$, $\bar B_c^* D_1^{(\prime)}$, $\bar B_c^* D_2^*$, $\bar B_{c0}^* D^*$, $\bar B_{c1}^{(\prime)} D^{(*)}$, $\bar B_{c2}^* D^*$ \\
& $1^{+}$ & $\eta_c \bar B^*$, $J/\psi \bar B^{(*)}$, $\bar B_c D^*$, $\bar B_c^* D^{(*)}$ & $\eta_c \bar B_0^*$, $\eta_c \bar B_1^{(\prime)}$, $\eta_c \bar B_2^*$, $J/\psi \bar B_0^*$, $J/\psi \bar B_1^{(\prime)}$, $J/\psi \bar B_2^*$, $h_c \bar B^{(*)}$, $\chi_{c0,1,2} \bar B^{(*)}$, $\bar B_c D_0^*$, $\bar B_c D_1^{(\prime)}$, $\bar B_c D_2^*$, $\bar B_c^* D_0^*$, $\bar B_c^* D_1^{(\prime)}$, $\bar B_c^* D_2^*$, $\bar B_{c0}^* D^{(*)}$, $\bar B_{c1}^{(\prime)} D^{(*)}$, $\bar B_{c2}^* D^{(*)}$ \\
& $2^{+}$ & $J/\psi \bar B^*$, $\bar B_c^* D^*$ & $\eta_c \bar B_1^{(\prime)}$, $\eta_c \bar B_2^*$, $J/\psi \bar B_0^*$, $J/\psi \bar B_1^{(\prime)}$, $J/\psi \bar B_2^*$, $h_c \bar B^{(*)}$, $\chi_{c0,1,2} \bar B^{(*)}$, $\bar B_c D_1^{(\prime)}$, $\bar B_c D_2^*$, $\bar B_c^* D_0^*$, $\bar B_c^* D_1^{(\prime)}$, $\bar B_c^* D_2^*$, $\bar B_{c0}^* D^{(*)}$, $\bar B_{c1}^{(\prime)} D^{(*)}$, $\bar B_{c2}^* D^{(*)}$ \\\hline
 $cb\bar c \bar s$ & $0^{+}$ & $\eta_c \bar B_s$, $J/\psi \bar B_s^*$, $\bar B_c D_s$, $\bar B_c^* D_s^*$ & $\eta_c \bar B_{s1}^{(\prime)}$, $J/\psi \bar B_{s0}^*$, $J/\psi \bar B_{s1}^{(\prime)}$, $J/\psi \bar B_{s2}^*$, $h_c \bar B_s^{(*)}$, $\chi_{c0} \bar B_s^*$, $\chi_{c1} \bar B_s^{(*)}$, $\chi_{c2} \bar B_s^*$, $\bar B_c D_{s1}^{(\prime)}$, $\bar B_c^* D_{s0}^*$, $\bar B_c^* D_{s1}^{(\prime)}$, $\bar B_c^* D_{s2}^*$, $\bar B_{c0}^* D_s^*$, $\bar B_{c1}^{(\prime)} D_s^{(*)}$, $\bar B_{c2}^* D_s^*$ \\
& $1^{+}$ & $\eta_c \bar B_s^*$, $J/\psi \bar B_s^{(*)}$, $\bar B_c D_s^*$, $\bar B_c^* D_s^{(*)}$ & $\eta_c \bar B_{s0}^*$, $\eta_c \bar B_{s1}^{(\prime)}$, $\eta_c \bar B_{s2}^*$, $J/\psi \bar B_{s0}^*$, $J/\psi \bar B_{s1}^{(\prime)}$, $J/\psi \bar B_{s2}^*$, $h_c \bar B_s^{(*)}$, $\chi_{c0,1,2} \bar B_s^{(*)}$, $\bar B_c D_{s0}^*$, $\bar B_c D_{s1}^{(\prime)}$, $\bar B_c D_{s2}^*$, $\bar B_c^* D_{s0}^*$, $\bar B_c^* D_{s1}^{(\prime)}$, $\bar B_c^* D_{s2}^*$, $\bar B_{c0}^* D_s^{(*)}$, $\bar B_{c1}^{(\prime)} D_s^{(*)}$, $\bar B_{c2}^* D_s^{(*)}$ \\
& $2^{+}$ & $J/\psi \bar B_s^*$, $\bar B_c^* D_s^*$ & $\eta_c \bar B_{s1}^{(\prime)}$, $\eta_c \bar B_{s2}^*$, $J/\psi \bar B_{s0}^*$, $J/\psi \bar B_{s1}^{(\prime)}$, $J/\psi \bar B_{s2}^*$, $h_c \bar B_s^{(*)}$, $\chi_{c0,1,2} \bar B_s^{(*)}$, $\bar B_c D_{s1}^{(\prime)}$, $\bar B_c D_{s2}^*$, $\bar B_c^* D_{s0}^*$, $\bar B_c^* D_{s1}^{(\prime)}$, $\bar B_c^* D_{s2}^*$, $\bar B_{c0}^* D_s^{(*)}$, $\bar B_{c1}^{(\prime)} D_s^{(*)}$, $\bar B_{c2}^* D_s^{(*)}$ \\\hline
 $cb\bar b \bar n$  & $0^{+}$ & $\eta_b D$, $\Upsilon D^*$, $B_c \bar B$, $B_c^* \bar B^*$ & $\eta_b D_1^{(\prime)}$, $\Upsilon D_0^*$, $\Upsilon D_1^{(\prime)}$, $\Upsilon D_2^*$, $h_b D^{(*)}$, $\chi_{b0} D^*$, $\chi_{b1} D^{(*)}$, $\chi_{b2} D^*$, $B_c \bar B_1^{(\prime)}$, $B_c^* \bar B_0^*$, $B_c^* \bar B_1^{(\prime)}$, $B_c^* \bar B_2^*$, $B_{c0}^* \bar B^*$, $B_{c1}^{(\prime)} \bar B^{(*)}$, $B_{c2}^* \bar B^*$ \\
& $1^{+}$ & $\eta_b D^*$, $\Upsilon D^{(*)}$, $B_c \bar B^*$, $B_c^* \bar B^{(*)}$ & $\eta_b D_0^*$, $\eta_b D_1^{(\prime)}$, $\eta_b D_2^*$, $\Upsilon D_0^*$, $\Upsilon D_1^{(\prime)}$, $\Upsilon D_2^*$, $h_b D^{(*)}$, $\chi_{b0,1,2} D^{(*)}$, $B_c \bar B_0^*$, $B_c \bar B_1^{(\prime)}$, $B_c \bar B_2^*$, $B_c^* \bar B_0^*$, $B_c^* \bar B_1^{(\prime)}$, $B_c^* \bar B_2^*$, $B_{c0}^* \bar B^{(*)}$, $B_{c1}^{(\prime)} \bar B^{(*)}$, $B_{c2}^* \bar B^{(*)}$ \\
& $2^{+}$ & $\Upsilon D^*$, $B_c^* \bar B^*$ & $\eta_b D_1^{(\prime)}$, $\eta_b D_2^*$, $\Upsilon D_0^*$, $\Upsilon D_1^{(\prime)}$, $\Upsilon D_2^*$, $h_b D^{(*)}$, $\chi_{b0,1,2} D^{(*)}$, $B_c \bar B_1^{(\prime)}$, $B_c \bar B_2^*$, $B_c^* \bar B_0^*$, $B_c^* \bar B_1^{(\prime)}$, $B_c^* \bar B_2^*$, $B_{c0}^* \bar B^{(*)}$, $B_{c1}^{(\prime)} \bar B^{(*)}$, $B_{c2}^* \bar B^{(*)}$ \\\hline
 $cb\bar b \bar s$  & $0^{+}$ & $\eta_b D_s$, $\Upsilon D_s^*$, $B_c \bar B_s$, $B_c^* \bar B_s^*$ & $\eta_b D_{s1}^{(\prime)}$, $\Upsilon D_{s0}^*$, $\Upsilon D_{s1}^{(\prime)}$, $\Upsilon D_{s2}^*$, $h_b D_s^{(*)}$, $\chi_{b0} D_s^*$, $\chi_{b1} D_s^{(*)}$, $\chi_{b2} D_s^*$, $B_c \bar B_{s1}^{(\prime)}$, $B_c^* \bar B_{s0}^*$, $B_c^* \bar B_{s1}^{(\prime)}$, $B_c^* \bar B_{s2}^*$, $B_{c0}^* \bar B_s^*$, $B_{c1}^{(\prime)} \bar B_s^{(*)}$, $B_{c2}^* \bar B_s^*$ \\
& $1^{+}$ & $\eta_b D_s^*$, $\Upsilon D_s^{(*)}$, $B_c \bar B_s^*$, $B_c^* \bar B_s^{(*)}$ & $\eta_b D_{s0}^*$, $\eta_b D_{s1}^{(\prime)}$, $\eta_b D_{s2}^*$, $\Upsilon D_{s0}^*$, $\Upsilon D_{s1}^{(\prime)}$, $\Upsilon D_{s2}^*$, $h_b D_s^{(*)}$, $\chi_{b0,1,2} D_s^{(*)}$, $B_c \bar B_{s0}^*$, $B_c \bar B_{s1}^{(\prime)}$, $B_c \bar B_{s2}^*$, $B_c^* \bar B_{s0}^*$, $B_c^* \bar B_{s1}^{(\prime)}$, $B_c^* \bar B_{s2}^*$, $B_{c0}^* \bar B_s^{(*)}$, $B_{c1}^{(\prime)} \bar B_s^{(*)}$, $B_{c2}^* \bar B_s^{(*)}$ \\
& $2^{+}$ & $\Upsilon D_s^*$, $B_c^* \bar B_s^*$ & $\eta_b D_{s1}^{(\prime)}$, $\eta_b D_{s2}^*$, $\Upsilon D_{s0}^*$, $\Upsilon D_{s1}^{(\prime)}$, $\Upsilon D_{s2}^*$, $h_b D_s^{(*)}$, $\chi_{b0,1,2} D_s^{(*)}$, $B_c \bar B_{s1}^{(\prime)}$, $B_c \bar B_{s2}^*$, $B_c^* \bar B_{s0}^*$, $B_c^* \bar B_{s1}^{(\prime)}$, $B_c^* \bar B_{s2}^*$, $B_{c0}^* \bar B_s^{(*)}$, $B_{c1}^{(\prime)} \bar B_s^{(*)}$, $B_{c2}^* \bar B_s^{(*)}$ \\
\hline\hline
\end{tabular*}
\end{center}
\end{table*}

The mass splittings for the $cb\bar c \bar n$, $cb\bar c \bar s$, $cb\bar b \bar n$, and $cb\bar b \bar s$ states are 274, 243, 333, and 309 MeV, respectively. Although all the mass spectra calculated here are significantly higher than those of color-magnetic interaction model~\cite{Chen:2016ont}, the mass splittings in these two studies coincide. However, the factors contributing to the fine structures are quite different in the color-magnetic interaction model and the extended relativized quark model. In the color-magnetic interaction model, only the spin-spin interactions appear, while in the extended relativized quark model, the Coulomb and confining potentials are also included. In addition to the spin-spin interactions, the Coulomb and confining potentials also contribute to mass splittings, owing to the nonzero off-diagonal matrix elements in the $cb\bar c \bar q$ and $cb\bar b \bar q$ systems. For instance, there are four configurations $|(cb)^{\bar 3}_0 (\bar c \bar n)^3_0 \rangle_0$, $|(cb)^{\bar 3}_1 (\bar c \bar n)^3_1\rangle_0$, $|(cb)^6_0 (\bar c \bar n)^{\bar 6}_0\rangle_0$, and $|(cb)^6_1 (\bar c \bar n)^{\bar 6}_1\rangle_0$ in the $J^P=0^+$ $cb \bar c \bar n$ systems. The Coulomb and confining potentials can lead to mixing effects between the configurations $|(cb)^{\bar 3}_0 (\bar c \bar n)^3_0 \rangle_0$ and $|(cb)^6_0 (\bar c \bar n)^{\bar 6}_0\rangle_0$, and also in the $|(cb)^{\bar 3}_1 (\bar c \bar n)^3_1 \rangle_0$ and $|(cb)^6_1 (\bar c \bar n)^{\bar 6}_1\rangle_0$. From Table~\ref{mass3}, it can be seen that the Coulomb and confining potentials play a more essential role than the spin-spin interactions in the fine structures.

It should be mentioned that the significant mass splittings originating from Coulomb and confining potentials for the mutiquarks are unique. In the traditional mesons and baryons, the color degrees of freedom are always trivial, and only spin-spin interactions contributes to the  mass splittings of $S-$wave ground states. The situation changes when we attempt to extend the pairwise potentials to the multiquark states. More complicated color degrees of freedom are involved, which can lead to extra contributions to the mass splittings. Further theoretical and experimental investigations are needed to test whether this generalization is reasonable or not.

\subsection{Discussions}

From the predicted mass spectra, we can see that all the triply-heavy tetraquarks lie above the corresponding meson-meson thresholds, which indicates that according to our model no stable $Q_1Q_2 \bar Q_3 \bar q_4$ state exists. In previous studies~\cite{Lu:2020rog,Lu:2020qmp}, we also investigated the mass spectra for the singly-heavy tetraquarks $Q_1q_2 \bar q_3 \bar q_4$ and doubly-heavy tetraquarks $Q_1Q_2 \bar q_3 \bar q_4$, and found that only the $IJ^P=01^+$ $bb \bar u \bar d$ state lay below the relevant strong and electromagnetic thresholds. In the case of the doubly-heavy tetraquarks $Q_1Q_2 \bar q_3 \bar q_4$, we pointed out that the mass ratio between the heavy $Q_1Q_2$ and light $\bar q_3 \bar q_4$ subsystems was the determining factor in the formation of a stable tetraquark~\cite{Lu:2020rog}. If one keeps reducing the mass of heavy quark $Q_2$, a doubly-heavy tetraquark $Q_1Q_2 \bar q_3 \bar q_4$ will become a singly-heavy one $Q_1q_2 \bar q_3 \bar q_4$. Similarly, replacing the $\bar q_3$ in a $Q_1Q_2 \bar q_3 \bar q_4$ state with $\bar Q_3$, one can obtain a triply-heavy tetraquark $Q_1Q_2 \bar Q_3 \bar q_4$. We can speculate that the empirical analysis of mass ratios for the doubly-heavy tetraquarks $Q_1Q_2 \bar q_3 \bar q_4$ also holds for the singly- and triply-heavy tetraquarks. Indeed, the stable $bb \bar u \bar d$ state has the largest mass ratio, while all other systems with smaller mass ratios cannot form bound compact tetraquarks. Caution should be exercised in generalizing this conclusion to the $Q_1Q_2 \bar Q_3 \bar Q_4$, $Q_1q_2 \bar Q_3 \bar q_4$ and $q_1q_2 \bar q_3 \bar q_4$ systems in which a charge conjugation quantum number may emerge, and then the above empirical conclusion may not hold. Moreover, it should be emphasized that, even if a tetraquark is not stable, it may subsist as a resonance with finite decay width and be observed by future experiments.

Besides the fall-apart decay mechanism, other two-body strong decay modes may also exist for the triply-heavy tetraquarks. As they have larger mass gaps, the higher $bb \bar c \bar n$, $cb \bar c \bar n$, and $cb \bar c \bar n$ states can decay into the lower ones by emitting pions. Certainly, like the $D_{s0}^*(2317)$ and $D_{s1}(2460)$ cases, the higher $cb \bar c \bar s$ and $cb \bar c \bar s$ states can also emit pions through the isospin violation process. Moreover, for the hidden charm and bottom states, the $c \bar c$ or $b \bar b$ quark pairs may first annihilate, and then these states can decay into the meson-meson or baryon-antibaryon final states as well as conventional heavy-light mesons.

\begin{figure}[!htbp]
\centering
\includegraphics[scale=0.9]{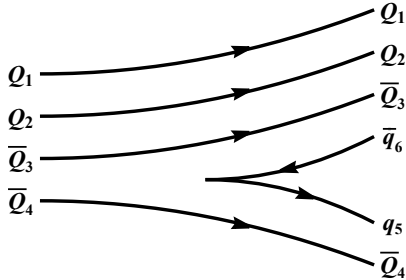}
\vspace{0.0cm} \caption{A fully-heavy tetraquark decays into the triply-heavy tetraquark plus heavy-light meson via light-quark pair creation.}
\label{qpc}
\end{figure}

Following the observation by the LHCb Collaboration of possible fully-heavy tetraquark structures in the $J/\psi J/\psi$ invariant mass, it is feasible to investigate the relation between fully- and triply-heavy tetraquarks. Indeed, the triply-heavy tetraquarks can be produced from the fully-heavy ones via light-quark pair creation; this decay process is shown in Fig.~\ref{qpc}. For the higher $cc\bar c \bar c$ tetraquarks, a typical decay chain is $cc\bar c \bar c \to cc\bar c \bar q + q \bar c \to c \bar c + c \bar q + q \bar c$, where the final sates can be $J/\psi D^{(*)} \bar D^{(*)}$ and $\eta_c D^{(*)} \bar D^{(*)}$. From our present calculations, the mass of the initial $cc\bar c \bar c$ tetraquarks should lie above 7256 MeV, and the triply-heavy tetraquarks $cc\bar c \bar q$ can be searched for in the $J/\psi D^{(*)}$ and $\eta_c D^{(*)}$ invariant masses by future experiments. Similar discussions can be generalized to other triply- and fully-heavy tetraquark systems.

Our present results are quite different from those of studies conducted within the color-magnetic interaction model~\cite{SilvestreBrac:1992mv,Cui:2006mp,Chen:2016ont} and QCD sum rule~\cite{Jiang:2017tdc}, and new contributions for the mass splittings from Coulomb and confining potentials emerge for the $cb\bar c \bar q$ and $cb\bar b \bar q$ systems. It is worth emphasizing that, to predict the tetraquarks, we adopt the extended relativized quark model and original parameters that describe the conventional mesons well and this unified treatment for mesons and tetraquarks is essential in order to obtain reliable mass spectra of triply-heavy tetraquarks. So far, it is difficult to distinguish these different models, owing to the lack of experimental information. Moreover, if different parameter schemes are adopted, significantly different results are obtained even in the color-magnetic interaction model~\cite{SilvestreBrac:1992mv,Cui:2006mp,Chen:2016ont}. Further investigations using various approaches and experimental searches are needed in order to clarify this problem.

\begin{figure}[!htbp]
\centering
\includegraphics[scale=0.9]{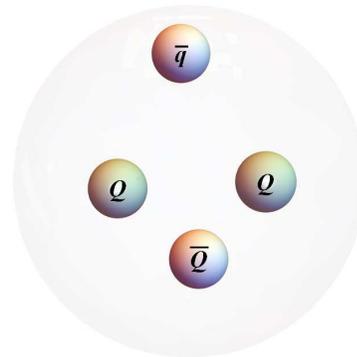}
\vspace{0.0cm} \caption{A typical sketch of the triply-heavy tetraquarks.}
\label{sketch}
\end{figure}

With the wave functions obtained, one can estimate the color proportions and expections of $\langle \boldsymbol r_{ij}^2  \rangle^{1/2}$ for the triply-heavy tetraquarks. Table~\ref{prop} shows these properties for the lowest states of various systems. Based on the relatively small radii, a typical sketch of the triply-heavy tetraquarks is shown in Fig.~\ref{sketch}. It can be seen that the four quarks are separated from on another in a compact tetraquark, which is significantly different from the diquark-antidiquark or loosely bound molecular picture. 

\begin{table*}[htp]
\begin{center}
\small
\caption{\label{prop} The color proportions and root mean square radii for the lowest states of various systems. The units of masses and $\langle \boldsymbol r_{ij}^2  \rangle^{1/2}$ are in MeV and fm, respectively.}
\begin{tabular*}{18cm}{@{\extracolsep{\fill}}*{10}{p{1.0cm}<{\centering}}}
\hline\hline
 System  & Mass  &   $|\bar 3 3\rangle$  &  $|6 \bar 6\rangle$  &    $\langle \boldsymbol r_{12}^2
 \rangle^{1/2}$    &  $\langle \boldsymbol r_{34}^2 \rangle^{1/2}$   &
 $\langle \boldsymbol r_{13}^2  \rangle^{1/2}$   &  $\langle \boldsymbol r_{24}^2  \rangle^{1/2}$
 & $\langle \boldsymbol r_{14}^2  \rangle^{1/2}$   &  $\langle \boldsymbol r_{23}^2  \rangle^{1/2}$ \\\hline
 $cc \bar c \bar n$  & 5400  &   70.1\%  &  29.9\%    &  0.434  &  0.502     &  0.403  &  0.595  &  0.595
 &  0.403 \\
 $cc \bar c \bar s$  & 5476  &   52.3\%  &  47.7\%   &  0.443  &  0.506     &  0.390  &  0.551  &  0.551
 &  0.390 \\
 $cc \bar b \bar n$  & 8658  &   47.2\%  &  52.8\%   &  0.440  &  0.487     &  0.353  &  0.585  &  0.585
 &  0.353 \\
 $cc \bar b \bar s$  & 8751  &   46.4\%  &  53.6\%   &  0.434  &  0.460     &  0.348  &  0.547  &  0.547
 &  0.348 \\
 $bb \bar c \bar n$  & 11873  &   97.8\%  &  2.2\%    &  0.280  &  0.478     &  0.351  &  0.545  &  0.545
 &  0.351 \\
 $bb \bar c \bar s$  & 11950  &   96.7\%  &  3.3\%    &  0.279  &  0.445     &  0.346  &  0.486  &  0.486
 &  0.346 \\
 $bb \bar b \bar n$  & 15107  &   50.4\%  &  49.6\%    &  0.305  &  0.482     &  0.274  &  0.536  &  0.536
 &  0.274 \\
 $bb \bar b \bar s$  & 15184  &   38.1\%  &  61.9\%    &  0.309  &  0.452     &  0.266  &  0.493  &  0.493
 &  0.266 \\
 $cb \bar c \bar n$  & 8560  &   49.5\%  &  50.5\%    &  0.382  &  0.517     &  0.412  &  0.547  &  0.611
 &  0.309 \\
 $cb \bar c \bar s$  & 8646  &   45.8\%  &  54.2\%    &  0.379  &  0.484     &  0.409  &  0.482  &  0.553
 &  0.307 \\
 $cb \bar b \bar n$  & 11735  &   37.8\%  &  62.6\%    &  0.376  &  0.479     &  0.366  &  0.522  &  0.587
 &  0.250 \\
 $cb \bar b \bar s$  & 11833  &   37.2\%  &  62.8\%    &  0.371  &  0.448     &  0.361  &  0.480  &  0.548
 &  0.246 \\
\hline\hline
\end{tabular*}
\end{center}
\end{table*} 

\begin{figure*}[!htb]
\centering
\includegraphics[scale=0.46]{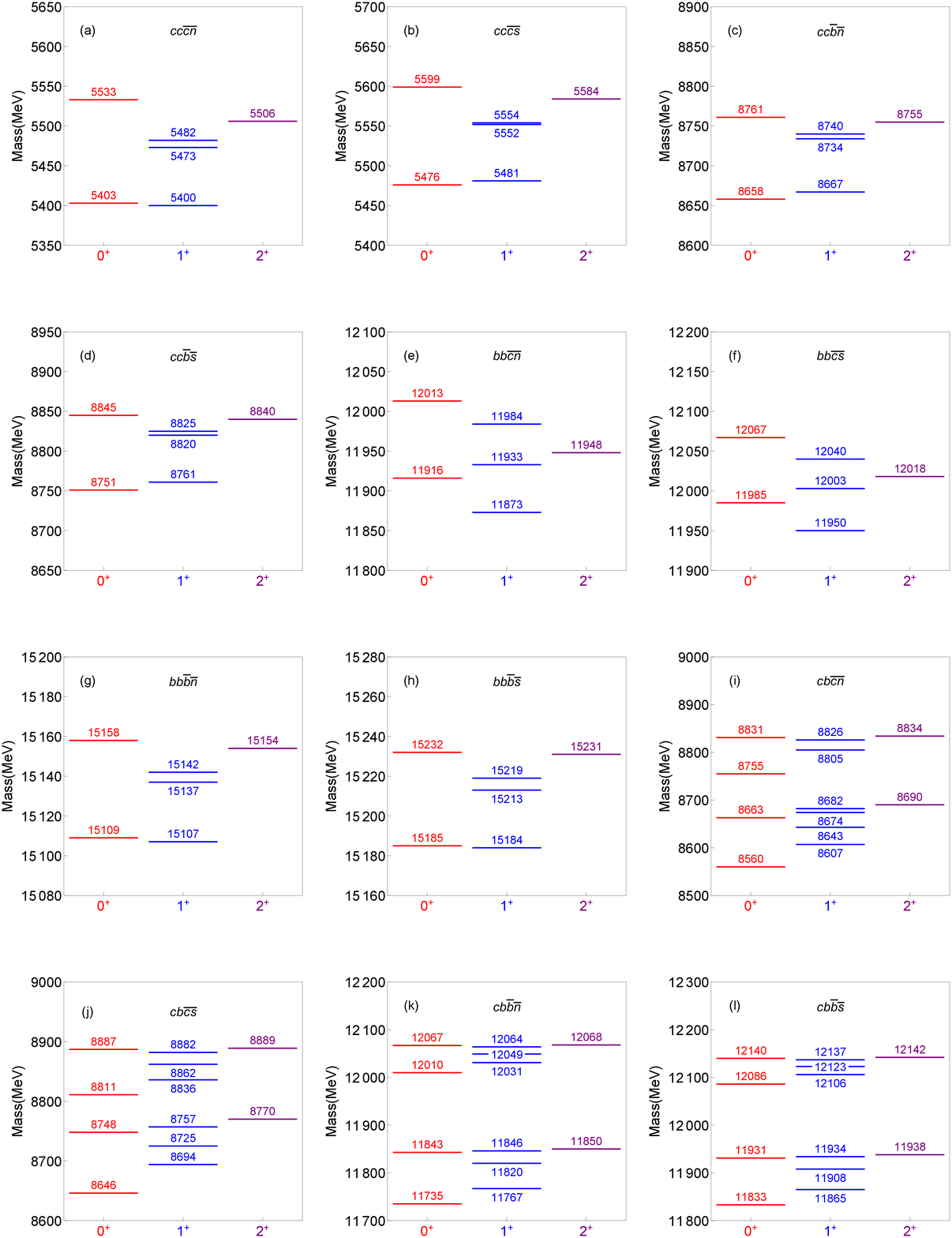}
\vspace{-0.0cm} \caption{The predicted mass spectra of triply heavy tetraquarks. (a)-(l) correspond to the mass spectra of $cc\bar c \bar n$, $cc\bar c \bar s$, $cc\bar b \bar n$, $cc\bar b \bar s$, $bb\bar c \bar n$, $bb\bar c \bar s$, $bb\bar b \bar n$, $bb\bar b \bar s$, $cb\bar c \bar n$, $cb\bar c \bar s$, $cb\bar b \bar n$, and $cb\bar b \bar s$ states, respectively.}
\label{mass}
\end{figure*}

The full mass spectra of $S-$wave triply-heavy tetraquarks are plotted in Figure~\ref{mass}. These similar mass patterns indicate that the approximate light flavor SU(3) symmetry and heavy quark symmetry are well preserved in these systems. So far, these two symmetries have been successfully applied to the conventional hadrons, and also appear in the case of the singly- and doubly-heavy tetraquarks. Definitely, they will also provide us with a powerful tool for investigating the triply-heavy tetraquarks. Unlike the extensive investigations into conventional heavy-light mesons and singly-heavy baryons, studies on triply-heavy tetraquarks are scarce, and we hope that more theoretical and experimental efforts will be made to increase our understanding of these heavy-light systems.

\section{Summary}{\label{Summary}}

In this paper, we use the relativized quark model to investigate the triply-heavy tetraquarks systematically. The mass spectra are obtained by solving the four-body relativized Hamiltonian including the Coulomb potential, confining potential, spin-spin interactions, and relativistic corrections. All of the triply-heavy tetraquarks are seen to lie above the corresponding meson-meson thresholds, and thus no stable binding one exists. The predicted mass splittings for the $cc\bar c \bar q$, $cc\bar b \bar q$, $bb\bar c \bar q$, and $bb\bar b \bar q$ systems are relatively small, while the mass gaps of the $cb\bar c \bar q$ and $cb\bar b \bar q$ states are significant. Moreover, in the $cb\bar c \bar q$ and $cb\bar b \bar q$ systems, the Coulomb and confining potentials also contribute to the mass splittings; this reveals unique features of the mutiquark states, which can be tested by further theoretical and experimental investigations.

As in the case of other heavy-light systems, it can be seen that all the mass spectra for triply-heavy tetraquarks show quite similar patterns, which preserve the light flavor SU(3) symmetry and heavy quark symmetry well. Also, it is worth emphasizing that, to predict the tetraquarks, we adopt the extended relativized quark model and original parameters that describe the conventional mesons well.

Through the fall-apart mechanism, these triply-heavy tetraquarks may easily decay into the heavy quarkonium plus heavy-light mesons. Also, some higher tetraquark states may emit pions into the lower ones, and the hidden charm and bottom states can decay into the meson-meson or baryon-antibaryon final states as well as the traditional heavy-light mesons via $c \bar c$ or $b \bar b$ quark pair annihilation. Further calculations on strong decay behaviors will be interesting and important for our understanding of these triply-heavy tetraquarks. We hope our predictions of triply-heavy tetraquarks can provide helpful information for future experimental searches.

\bigskip
\noindent
\begin{center}
	{\bf ACKNOWLEDGEMENTS}\\
\end{center}

We would like to thank Xian-Hui Zhong for his valuable discussions. 
This work is supported by the National Natural Science Foundation of China under Grants No.~11705056, No.~11775050, No.~11947224,
No.~11475192, No.~11975245, and No.~U1832173. This work is also supported by the State Scholarship Fund of China Scholarship Council under Grant No. 202006725011, the Sino-German CRC 110 ``Symmetries and the Emergence of Structure in QCD'' project by NSFC under the Grant No.~12070131001, the Key Research Program of Frontier Sciences, CAS, under the Grant No.~Y7292610K1, and the National Key Research and Development Program of China under Contracts No. 2020YFA0406300.

\end{document}